\documentclass[11pt]{article}
\usepackage{amssymb,amsmath,epsfig,graphics,mathabx}
\usepackage{mathtools}
\usepackage{enumerate}
\usepackage{braket}
\usepackage{mathtools}
\usepackage{mathabx}
\newsavebox{\accentbox}
\providecommand{\abs}[1]{\lvert#1\rvert}

\usepackage[top=2cm, bottom=2.5cm, left=2.5cm, right=2.5cm]{geometry}

\providecommand{\abs}[1]{\lvert#1\rvert}

\usepackage{graphicx,amssymb,amsmath,amsthm,amsfonts,mathtools,makeidx,bm,enumitem,float,relsize,empheq,lscape,emptypage,caption,bm,sectsty,xparse,cancel,mathrsfs,pifont,lipsum,tikz,soul,xcolor}
\usepackage[framemethod=tikz]{mdframed}
\usepackage[most]{tcolorbox}
\usepackage{subfigure}
\numberwithin{equation}{section}
\usepackage{xparse}
\usepackage{braket}
\usepackage{physics}
\usepackage{cite}
\usepackage{titlesec}

\subsubsectionfont{\color{black}}

\usepackage{mathpazo}
\ExplSyntaxOn
\NewDocumentCommand{\MeijerG}{smmmm}
 {
  \IfBooleanTF{#1}
   {
    \vic_meijerg:nnnnnn { #2 } { #3 } { #4 } { #5 } { small } { }
   }
   {
    \vic_meijerg:nnnnnn { #2 } { #3 } { #4 } { #5 } { } { \; }
   }
 }

\seq_new:N \l__vic_meijerg_args_in_seq
\seq_new:N \l__vic_meijerg_args_out_seq

\cs_new_protected:Nn \vic_meijerg:nnnnnn
 {
  \seq_set_split:Nnn \l__vic_meijerg_args_in_seq { | } { #3 }
  \seq_clear:N \l__vic_meijerg_args_out_seq  
  \seq_map_inline:Nn \l__vic_meijerg_args_in_seq
   {
    \seq_put_right:Nn \l__vic_meijerg_args_out_seq
     {
      \begin{#5matrix} ##1 \end{#5matrix}
     }
   }
  G\sp{#1}\sb{#2}
  \left(
  \seq_use:Nn \l__vic_meijerg_args_out_seq { #6\middle|#6 }
  #6\middle|#6
  #4
  \right)
 }
\ExplSyntaxOff

\begin{document}
\begin{sloppypar}
\vspace*{0cm}
\begin{center}
{\setlength{\baselineskip}{1.0cm}{ {\Large{\bf 
EQUIVALENT NON-RATIONAL EXTENSIONS OF THE HARMONIC OSCILLATOR, THEIR LADDER OPERATORS AND COHERENT STATES
 \\}} }}
\vspace*{1.0cm}
{\large{\sc{Alonso Contreras-Astorga}$^{1,2,\dagger}$},\quad {\sc{David J. Fernández C.}$^{3,\ddagger}$},\quad and\quad {\sc{César Muro-Cabral}$^{3,4,*}$}}
\indent \vspace{.3cm} 
\noindent \\	
$~^1$ CONACyT - Departamento de Física, Cinvestav, A.P. 14-740, 07000 México D.F., Mexico\\
$~^2$ Canadian Quantum Research Center,  204-3002 32 Ave Vernon, BC V1T 2L7, Canada\\
$~^3$ Departamento de Física, Cinvestav, A.P. 14-740, 07000 México D.F., Mexico\\
$~^4$ Unidad Querétaro, Cinvestav, 76230, Querétaro,
Mexico. \\  \vspace{.3cm}
$^\dagger$E-mail:
alonso.contreras@conacyt.mx 
\\ ${}^\ddagger$E-mail:
david@fis.cinvestav.mx  
\\ ${}^*$E-mail:
cmuro@fis.cinvestav.mx 
\end{center}

\vspace*{.5cm}
\begin{abstract}
\noindent
 In this work, we generate a family of quantum potentials that are non-rational extensions of the harmonic oscillator. Such a family can be obtained via two different but equivalent supersymmetric transformations. We construct ladder operators for these extensions as the product of the intertwining operators of both transformations. Then, we generate families of Barut-Girardello coherent states  and analyze some of their properties as temporal stability, continuity on the label, and completeness relation. Moreover, we calculate mean-energy values, time-dependent probability densities, Wigner functions, and the Mandel Q-parameter to uncover a general non-classical behavior of these states. 
\end{abstract}


\section{Introduction}

In quantum mechanics, there has always been a continuing interest in  finding exactly solvable potentials. Supersymmetric quantum mechanics (SUSY), closely related to the factorization method or intertwining technique, has been an efficient tool to construct solvable quantum  potentials departing from a given initial one \cite{matveev92,Khare95,Bagchi01,Andrianov04,david2010supersymmetric,gangopadhyaya17,junker19}. For years, this technique has been attractive to study a wide range of systems on physics and mathematical-physics, e.g., the interactions between Dirac Fermions and electromagnetic fields \cite{Kuru2009,Midya2014,castillo2020dirac,contreras2020super,daniel2020electron,Bagchi2021}, optical systems  \cite{Miri2013a,contreras2019photonic,Chandra2021}, generation of solutions for non-linear equations \cite{matveev92,Adler1994,bermudez2011supersymmetric,Correa2016,clarkson2020cyclic}, superintegrable systems \cite{Demircioglu2002,Marquette2009}, etc. In addition, for some potentials the state-deleting (Krein-Adler) \cite{adler1994modification} and state-adding (Darboux-Crum)  \cite{gomez2014extended} SUSY transformations  have received special attention, since the eigenfunctions of the rational extensions are written in terms of exceptional orthogonal polynomials, which are extensions of the classical Hermite, Laguerre and Jacobi polynomials \cite{Gomez-Ullate2009,Quesne2008,Odake2009}. Furthermore, in some cases there is an equivalence between the Krein-Adler and Darboux-Crum transformations which has been suitable for generating ladder operators for these rational extensions \cite{marquette2013new,marquette2014combined} and supplied a new path to study superintegrable systems \cite{marquette2020fourth}, to generate solutions for the Painlevé equations \cite{clarkson2020cyclic,marquette2022family}, etc. 
Let us point out that all these previous works have been done taking into account only supersymmetric rational extensions. As the first contribution of this work we will show that a similar equivalence between two SUSY transformations can be found by employing not only polynomial seed solutions but general seed solutions, giving place to non-rational extensions of the harmonic oscillator.


 On the other hand, it is well-known that Schr\"odinger found for the first time the coherent states of the harmonic oscillator in 1926  \cite{schrodinger1926stetige}. Later on, in his seminal work Glauber rediscovered these states when giving a quantum description of coherent light \cite{glauber1963coherent}. As a consequence, coherent states became an important research field in quantum physics, since they reproduce the classical behavior in phase space. Furthermore, the study and derivation of coherent states for other quantum systems have become an intensive research subject because they allow examining in phase-space the system behavior at the border between quantum and classical regimes \cite{barut71,perelomov72,nieto79,dodonov80,gazeau1999coherent}.  A subject running parallelly to SUSY is the coherent states construction for potentials generated through such a technique \cite{hussin1994coherent,Kumar96,Junker99,Bagchi99,fernandez2007coherent}.

The coherent states of the harmonic oscillator are given by \cite{glauber1963coherent}
\begin{equation}
    \ket{z}=e^{-\frac{\abs{z}^{2}}{2}}\sum^{\infty}_{n=0}\frac{z^{n}}{\sqrt{n!}}\ket{n},
\end{equation}
where $z\in \mathbb{C}$, and $\ket{n}$  are the number states of the harmonic oscillator with energy eigenvalues $E_{n}=n+1/2$. These quantum states can be derived from four definitions (the first three were introduced by Glauber whereas the fourth was proposed by Gazeau and Klauder \cite{gazeau1999coherent}):
\begin{enumerate}
    \item  They are eigenstates of the annihilaton operator,
    \begin{equation}\label{definition1}
        a^{-}\ket{z}=z\ket{z}.
    \end{equation}
\item They are obtained by applying the displacement operator $D(z)\equiv e^{z a^{+}-z^{*}a^{-}}$ on the harmonic oscillator vacuum state, i.e.
\begin{equation}\label{dispop}
    \ket{z}=D(z)\ket{0}.
\end{equation}

\item They are states with minimum uncertainty relation for the position $q$ and momentum $p$ operators: 
\begin{equation}
    \langle \bigtriangleup q \rangle^{2} \langle \bigtriangleup p \rangle^{2}=\frac{1}{4}.
\end{equation}
\item They are an overcomplete set satisfying the completeness relation
\begin{equation} \label{fourth def}
       \frac{1}{\pi} \int_{\mathbb{C}}\ket{z}\bra{z}d^{2}z=\mathbb{1}_{\mathbb{H}}.
    \end{equation}
\end{enumerate}

These definitions, all of them based on properties of the harmonic oscillator coherent states, have been employed to construct coherent states for different quantum systems \cite{gazeau1999coherent}. When the coherent states are constructed using the first two definitions, the first task is to find suitable ladder and/or displacement operators. It is important to stress that, in general, these definitions lead to different families of coherent states.

In previous works, it has been derived coherent states for supersymmetric extensions of several interesting potentials, as the potential well, the harmonic and truncated oscillators, the Rosen-Morse potentials and others \cite{hussin1994coherent,hussin1999higher,fernandez2007coherent,Bermudez2014,hussin2017coherent,hoffmann2018coherent,hoffmann2018non,hussin2019coherent,hoffmann2019ladder,garneau2021ladder}. In this is article, we are interested in deriving  coherent states as eigenstates of the annihilation operator (also called Barut-Girardello coherent states \cite{barut71}) for non-rational extensions of the harmonic oscillator by building the ladder operators as a product of the intertwining operators related to two equivalent transformations, i.e.  through the scheme proposed in \cite{marquette2013new,marquette2014combined}.

The organization of this paper is as follows: in the next two Sections we introduce the SUSY transformations and define the polynomial Heisenberg algebras. In the fourth Section we summarize the basis of the equivalent rational extensions of the harmonic oscillator and the ladder operators that can be built from these extensions. In the fifth Section we present a family of SUSY equivalent non-rational extensions that can be generated by using general solutions of the stationary Schr\"odinger equation and the ladder operators associated to these extensions. In  Section 6 we derive the corresponding coherent states as eigenstates of the annihilation operator and analyze some properties, as  temporal stability, continuity on the label, completeness relation, mean energy values, time evolving probability densities, Wigner functions and Mandel-Q parameter. In the last section, we present our conclusions. 

\section{Supersymmetric quantum mechanics}
Let us suppose that $H$ is an initial solvable Hamiltonian with eigenfunctions $\psi_{n}(x)$ and eigenvalues $E_n$, $n=0,1,2,\dots$. It is assumed that the following intertwining relations are fulfilled
\begin{equation}
    \widetilde{H}B^{+}=B^{+}H,\quad HB=B\widetilde{H}.
\end{equation}
where $B,B^{+}$ are $k$th-order differential operators and $H,\widetilde{H}$ are the two Schr\"odinger Hamiltonians
\begin{equation}
    \widetilde{H}=-\frac{1}{2}\frac{d^{2}}{dx^{2}}+\widetilde{V}(x),\quad H=-\frac{1}{2}\frac{d^{2}}{dx^{2}}+V(x).
\end{equation}
We suppose as well that $k$ solutions of the  Schr\"odinger equation
\begin{equation}\label{stationaryeq}
    Hu_{j}=\epsilon_{j}u_{j}, \quad j=1,~2, \dots, k, 
\end{equation}
for $k$ different factorization energies $\epsilon_{j}$ are given, which are called seed solutions in the literature \cite{quesne2011higher}. Then, the SUSY partner potential $\widetilde{V}$ will be given by
\begin{equation} \label{susypoten}  
\widetilde{V}=V-[\ln{ W(u_{1},u_{2},\dots,u_{k}) }]'',
\end{equation}
where $W(f_{1},f_{2},\dots,f_{k})$ denotes the Wronskian of $f_1,~f_2, \dots,f_{k}$. In order to obtain a regular potential inside the domain $x$ of $V(x)$, the seed solutions must be chosen such that the Wronskian in \eqref{susypoten} has not zeroes in such a domain.

The eigenfunctions of the Hamiltonian $\widetilde{H}$ can be obtained via
\begin{equation}
    \widetilde{\psi}_{n}=\frac{B^{+}\psi_{n}}{\sqrt{(E_{n}-\epsilon_{1})\dots(E_{n}-\epsilon_{k})}} = \frac{1}{\sqrt{2^{k}(E_{n}-\epsilon_{1})\dots(E_{n}-\epsilon_{k})}} \frac{W(u_{1},u_{2},\dots,u_{k},\psi_{n})}{W(u_{1},u_{2},\dots,u_{k})}, 
\end{equation}
with eigenvalue $E_{n}$ where the states fulfilling $B^{+}\psi_{n}= 0$ are excluded. Furthermore, the SUSY partner Hamiltonian $\widetilde{H}$ could have a finite number of additional eigenfunctions $\widetilde{\psi}_{\epsilon_{j}}$, known as missing states, which corresponding eigenvalues are the factorization energies $\epsilon_{j}$. They can be written as
\begin{equation}
    \widetilde{\psi}_{\epsilon_{j}}\propto \frac{W(u_{1},\dots,u_{j-1},u_{j+1},\dots,u_{k})}{W(u_{1},u_{2},\dots,u_{k}) }. 
\end{equation}
Depending on the square integrability of such missing states, the factorization energies could belong to the spectrum of $\widetilde{H}$.  The intertwining operators $B^{+},B$ satisfy as well the following factorizations: 
\begin{align}
    B^{+}B=(\widetilde{H}-\epsilon_{1})\dots(\widetilde{H}-\epsilon_{k}), \quad
    BB^{+}=(H-\epsilon_{1})\dots(H-\epsilon_{k}).
\end{align}

As a historical remark, let us note that it is possible to realize the Witten supersymmetric algebra with two generators $[Q_{i},H^{ss}]=0, ~ \{ Q_{i},Q_{j} \}=\delta_{ij}H^{ss},~ i,j=1,2,$ by defining \cite{witten1981dynamical}:
\begin{align}
  & Q=\left(\begin{array}{cc}
0 & 0 \\
B & 0
\end{array} \right),\quad Q^{+}=\left(\begin{array}{cc}
0 & B^{+} \\
0 & 0
\end{array} \right), \quad H^{ss}=\left(\begin{array}{cc}
B^{+}B & 0 \\
0 & BB^{+}
\end{array} \right), \nonumber \\
&  Q_{1}=\frac{Q^{+}+Q}{\sqrt{2}},\quad Q_{2}=\frac{Q^{+}-Q}{i\sqrt{2}}, 
\end{align}
where $H^{ss}$ is called supersymmetric Hamiltonian.

\subsection{Second-order supersymmetric quantum mechanics}

It is of particular interest for this paper the second-order SUSY, thus, let us briefly review this case (for more details see \cite{david2010supersymmetric}). We depart from the intertwining relation $\widetilde{H} B^+= B^+ H$, where $B^{+}$ is a second-order differential operator:
\begin{equation} \label{B+}
B^{+}=\frac{1}{2}\left[\frac{d^{2}}{dx^{2}}-g(x)\frac{d}{dx}+h(x)\right].    
\end{equation} 
The functions $g(x),h(x)$ can be expressed in terms of the two seed solutions $u_{1},u_{2}$ with corresponding factorization energies $\epsilon_{1},\epsilon_{2}$, as follows:  
\begin{align}\label{gfuncrb}
 g=\frac{W'(u_{1},u_{2})}{W(u_{1},u_{2})}, \quad  h=\frac{g'}{2}+\frac{g^{2}}{2}-2V+\frac{\epsilon_1+ \epsilon_2}{2}. 
\end{align}
Therefore, the SUSY partner potential can be written as $\widetilde{V}=V-[\ln{W(u_{1},u_{2})}]''$.
The eigenfunctions of the Hamiltonian $\widetilde{H}$ are given by
\begin{equation}
\label{eigensecond}
    \widetilde{\psi}_{n}=\frac{B^{+}\psi_{n}}{\sqrt{(E_{n}-\epsilon_{1})(E_{n}-\epsilon_{2})}},
\end{equation}
while the missing states by
\begin{equation}
    \widetilde{\psi}_{\epsilon_{1}}\propto \frac{u_{2}}{W(u_{1},u_{2})},\quad \widetilde{\psi}_{\epsilon_{2}}\propto\frac{u_{1}}{W(u_{1},u_{2})}.
\end{equation}
As previously mentioned, the normalizability of these missing states will determine the energy spectrum of the Hamiltonian $\widetilde{H}$. Moreover, the operator $B=(B^+)^+$ becomes 
\begin{equation} \label{B-}
B=\frac{1}{2}\left[\frac{d^{2}}{dx^{2}}+g(x)\frac{d}{dx}+g'(x)+h(x)\right].     
\end{equation}
This operator links the eigenfunctions of $\widetilde{H}$ with those of $H$ in the way $\psi_n \propto B \widetilde{\psi}_n$, and it annihilates the missing states, $B \widetilde{\psi}_{\epsilon_j}=0$. Finally, the intertwining operators $B^+, B$ fulfill the two factorizations $ B^{+}B=(\widetilde{H}-\epsilon_{1})(\widetilde{H}-\epsilon_{2})$ and $BB^{+}=(H-\epsilon_{1})(H-\epsilon_{2})$.
\section{Polynomial Heisenberg algebras}

Let us take as $H$ the harmonic oscillator Hamiltonian,
\begin{equation}
    H=-\frac{1}{2}\frac{d^{2}}{dx^{2}}+\frac{1}{2}x^{2}, \label{HO H}
\end{equation}
whose eigenfunctions and eigenvalues are given by 
\begin{align}
    \psi_{n}(x)=\sqrt{\frac{1}{2^{n}\sqrt{\pi}n!}}e^{-\frac{x^{2}}{2}}H_{n}(x),\quad E_{n}=n+\frac{1}{2},\quad n=0,1,2,\dots,
\end{align}
where $H_{n}(x)$ is the Hermite polynomial of $n$th degree. The first-order differential ladder operators
\begin{equation} \label{a amas}
   a^{-}=\frac{1}{\sqrt{2}}\left(\frac{d}{dx}+x\right),\quad a^{+}=\frac{1}{\sqrt{2}}\left(-\frac{d}{dx}+x\right),
\end{equation}
factorize the oscillator Hamiltonian as $a^{+}a^{-}=H-\frac{1}{2}$ and $a^{-}a^{+}=H+\frac{1}{2}$;
they also generate the Lie algebra  (Heisenberg-Weyl) $[H,a^{\pm}]=\pm  a^{\pm},~ [a^{-},a^{+}]=\mathbb{1},$ which indicates that work as ladder operators.

The polynomial Heisenberg algebras (PHA) are deformations of the previous algebra, which are realized by a one dimensional Schrödinger Hamiltonian $\widetilde{H}$ and two finite-order differential ladder operators $\mathcal{L}^{\pm}$, such that $\mathcal{L}^{\pm}$ commute with $\widetilde{H}$ as the harmonic oscillator Hamiltonian does but between them commute as to produce a $m$-th degree polynomial $P_{m}(\widetilde{H})$ in $\widetilde{H}$ \cite{carballo2004polynomial}, i.e.,
\begin{align}
    [\widetilde{H},\mathcal{L}^{\pm}]=\pm \omega \mathcal{L}^{\pm}, \qquad 
    [\mathcal{L}^{-},\mathcal{L}^{+}]=N_{m+1}(\widetilde{H}+\omega)-N_{m+1}(\widetilde{H})=P_{m}(\widetilde{H}), \label{PHA}
\end{align}
with $\omega$ being a positive constant and
\begin{align}
    N_{m+1}(\widetilde{H})=\mathcal{L}^{+}\mathcal{L}^{-}=\displaystyle\prod_{i=1}^{m+1}(\widetilde{H}-\mathcal{E}_{i}). \label{Number}
\end{align}
To determine the energy spectrum of systems ruled by PHA, it is important to identify the formal eigenstates of $\widetilde{H}$ belonging as well to the kernel  $K_{\mathcal{L}^{-}}$ of the annihilation operator $\mathcal{L}^{-}$, known as extremal states. Since $K_{\mathcal{L}^{-}}$ is invariant under $\widetilde{H}$,
\begin{equation}
\mathcal{L^{-}}\psi =0, \quad \Rightarrow \quad  \mathcal{L^{+}}\mathcal{L^{-}}\psi = \displaystyle\prod_{i=1}^{m+1}(\widetilde{H}-\mathcal{E}_{i}) \psi = 0,  \quad \forall \, \psi \in  K_{\mathcal{L}^{-}},
\end{equation}
it is natural to take as a basis of $K_{\mathcal{L}^{-}}$ the states $\psi_{\mathcal{E}_{i}}$ such that $ \widetilde{H} \psi_{\mathcal{E}_{i}}=\mathcal{E}_{i}\psi_{\mathcal{E}_{i}}$.
Therefore, in principle we can generate $m+1$ energy ladders with spacing $\bigtriangleup E=\omega$ by applying the operator $\mathcal{L}^{+}$ onto each extremal state. However, if only $s$ extremal states satisfy the boundary conditions (known as physical extremal states), then from the iterated action of $\mathcal{L}^{+}$ we will obtain only $s$ physical energy ladders.

\section{Harmonic oscillator rational extensions and ladder operators}

A typical technique employed to generate rational extensions of the harmonic oscillator is the Krein-Adler (or state-deleting) transformation, which consists in taking as seed solutions eigenfunctions of the Hamiltonian, $\lbrace u_{i}^{KA}\rbrace_{i=1}^{k}=\lbrace \psi_{d_{i}} \rbrace_{i=1}^{k}$, in such a way that the Krein condition,
\begin{equation}
\displaystyle\prod_{i=1}^{k}(n-d_{i})\geq 0, \quad \forall \, n \in \mathcal{Z}^{+} , 
\end{equation}
is fulfilled. This means that the spectrum of the partner Hamiltonian $\widetilde{H}$  will contain only even gaps (which are composed by an even number of  missing consecutive levels) \cite{adler1994modification}.

On the other hand, by replacing $x\to ix$ in the harmonic oscillator Schr\"odinger equation $H \psi = E \psi$, we can find solutions for the negative energy parameters $E_{-(m+1)}=-\left(m+\frac{1}{2}\right)$, $m=0,1,2,\dots$, which are given by
\begin{equation}\label{vfunc}
    \varphi_{m}(x)=e^{\frac{x^{2}}{2}}(-i)^{m}H_{m}(ix). 
\end{equation}
They are nodeless for $m$ even and posses a single node at $x=0$ for $m$ odd. Moreover, for even $m$  the reciprocal of $\varphi_{m}(x)$ is square-integrable. Thus, another SUSY transformation used to generate rational extensions of the oscillator is by state-adding levels, employing as seeds the non-normalizable solutions (\ref{vfunc}). These kind of extensions are usually called Darboux-Crum transformations \cite{odake2013krein,gomez2014extended}.

It has been a subject of study to explore the equivalence between the Krein-Adler and Darboux-Crum transformations for generating equivalent rational extensions of shape-invariant potentials \cite{odake2013krein,gomez2013rational,marquette2013new,marquette2014combined,gomez2014extended}. Once a Krein-Adler transformation has been performed, as the one previously mentioned, then to get the same potential with a Darboux-Crum transformation, up to a global energy shift, it must be taken as seed solutions  $\lbrace u_{i}^{DC}\rbrace=\lbrace \varphi_{1},\dots,\Breve{\varphi}_{d_{k}-d_{k-1}},\Breve{\varphi}_{d_{k}-d_{k-2}},\dots,\Breve{\varphi}_{d_{k}-d_{1}},\varphi_{d_{k}} \rbrace$, where the symbol $\Breve{\varphi}_{i}$ indicates that the solution $\varphi_{i}$ is excluded from the set. As a result, it is found that the two SUSY generated potentials are related by \cite{odake2013krein,gomez2014extended}
\begin{equation}
    \widetilde{V}^{KA}(x)=\widetilde{V}^{DC}(x)+d_{k}+1.
\end{equation}
Then, a necessary but not sufficient condition to produce SUSY equivalent partner Hamiltonians $\widetilde{H}^{KA},~\widetilde{H}^{DC}$ is to ask that $\text{Sp}[\widetilde{H}^{KA}]=\text{Sp}[\widetilde{H}^{DC}+d_k+1]$. 

It is important to look for ladder operator that connect the energy levels of the SUSY partner Hamiltonians. The most known ladder operators for the SUSY extensions of the harmonic oscillator \cite{mielnik1984factorization,hussin1999higher} are written as  $ \mathcal{L}^{-}=B^{+}a^{-}B,~ \mathcal{L}^{+}=B^{+}a^{+}B$, 
where $B,~B^{+}$ are the intertwining operators of the SUSY transformation while $a^{-},~a^{+}$ are the standard ladder operators of the oscillator given in equation \eqref{a amas}. These  operators connect the energy levels that are in the part of the spectrum that is isospectral to the harmonic oscillator, but the missing states are completely disconnected since $B  {\widetilde\psi}_{\epsilon_j}=0$. Furthermore, in Ref. \cite{Bermudez2014} there were found general third-order ladder operators that behave as the natural ladder operators but allow to connect as well the new equidistant energy levels generated below the harmonic oscillator ground state. However, both subspaces, associated to the isospectral part and to the new energy levels, still remain disconnected.

An alternative way to build a set of ladder operators for the SUSY rational extensions of the harmonic oscillator was introduced in \cite{marquette2013new,marquette2014combined}. Let us suppose that $B_{KA},B_{DC}$ are the intertwining operators related to the equivalent Krein-Adler and Darboux-Crum extensions, respectively. Thus, if we define the operators
\begin{equation}\label{ladderm}
    \mathcal{L}^{-}=B_{KA}^{+}B_{DC},\quad \mathcal{L}^{+}=B_{DC}^{+}B_{KA},
\end{equation}
they fulfill the polynomial Heisenberg algebra
\begin{align}
    [\widetilde{H}^{DC},\mathcal{L}^{\pm}]=(d_{k}+1)\mathcal{L}^{\pm},\qquad 
    [\mathcal{L}^{-},\mathcal{L}^{+}]=N_{d_{k}+1}(\widetilde{H}^{DC}+d_{k}+1)-N_{d_{k}+1}(\widetilde{H}^{DC}).
\end{align}
Therefore, these ladder operators generate ladders with energy spacing length of $d_{k}+1$. 
The coherent states of these rational extensions of the harmonic oscillator were derived straightforwardly as eigenstates of the annihilation operator $\mathcal{L}^{-}$ of \eqref{ladderm} in  \cite{hoffmann2018coherent,hoffmann2019ladder}. 

\section{Equivalent non-rational extensions of the harmonic oscillator and ladder operators}

In order to generate equivalent non-rational extensions, let us first write down the general solution of the stationary Schr\"odinger equation (\ref{stationaryeq})  for the harmonic oscillator Hamiltonian \eqref{HO H} with energy parameter $\mathcal{E}= \lambda + 1/2$, as follows 
\begin{equation}
    u(x)=e^{-\frac{x^2}{2}}[H_{\lambda}(x)+\gamma H_{\lambda}(-x)],
\end{equation}
where 
\begin{equation}\label{hermitefunc}
    H_{\lambda}(x)\equiv \frac{2^{\lambda}\Gamma\left(\frac{1}{2}\right)}{\Gamma\left(\frac{1-\lambda}{2} \right)}{}_{1}F_{1}\left(-\frac{\lambda}{2};\frac{1}{2};x^{2}\right)+ \frac{2^{\lambda} \Gamma\left(-\frac{1}{2}\right) }{\Gamma\left(-\frac{\lambda}{2} \right)} x {}_{1}F_{1}\left(\frac{1-\lambda}{2};\frac{3}{2};x^{2}\right), 
\end{equation}
are the Hermite functions \cite{weisner1959generating},
\begin{equation}
    _{1}F_{1}(a;b;z)=\frac{\Gamma(b)}{\Gamma(a)}\sum_{n=0}^{\infty}\frac{\Gamma(a+n)}{\Gamma(b+n)}\frac{z^{n}}{n!},
\end{equation}
is the confluent hypergeometric function, 
and $\gamma$ is a real parameter. If $\gamma>0$, the solution $u(x)$ will have an even number of zeroes while for $\gamma<0$ this number will be odd. The solution $u(x)$ is closely related as well to the parabolic cylinder functions \cite{abramowitz}.


Let us generate now a SUSY partner potential $V^{(1)}$ of the harmonic oscillator through a second-order transformation. We employ the seed solutions 
\begin{equation}
    u_{1}^{(1)}(x)=e^{-\frac{x^2}{2}}[H_{\lambda_{1}}(x)+\gamma H_{\lambda_{1}}(-x)],\quad u_{2}^{(1)}(x)=\varphi_{1}(x),
\end{equation}
with factorization energies $-3/2 <\mathcal{E}_{1}<1/2$, and $\mathcal{E}_{2}=E_{-2}=-3/2$, respectively with $\lambda_1= \mathcal{E}_{1}-1/2$. To obtain a nodeless Wronskian $W(u_{1}^{(1)},u_{2}^{(1)})$ we must take $\gamma >0$. From equation \eqref{susypoten}, the SUSY partner potential $V^{(1)}$ is
\begin{equation}
    V^{(1)}=\frac{x^{2}}{2}-[\ln W(u_{1}^{(1)},u_{2}^{(1)})]''. 
\end{equation}
The explicit expressions of the second-order intertwining operators $B^{(1)},B^{(1)+}$ can be obtained from equations \eqref{B+}, and \eqref{B-}. They  relate the Hamiltonian $H^{(1)}$ with the oscillator Hamiltonian (\ref{HO H}) in the way
\begin{equation}
    H^{(1)}B^{(1)+}=B^{(1)+}H, \quad H B^{(1)}=B^{(1)}H^{(1)}.
\end{equation}
The eigenfunctions of this Hamiltonian are
\begin{equation}
    \psi_{n}^{(1)}=\frac{B^{(1)+}\psi_{n}}{\sqrt{(E_{n}-\mathcal{E}_{1})(E_{n}-\mathcal{E}_{2})}},\quad n=0,1,2,\dots,
\end{equation}
and the missing states read
\begin{equation}
    \psi^{(1)}_{\mathcal{E}_{1}}\propto\frac{u_{2}^{(1)}}{W(u_{1}^{(1)},u_{2}^{(1)})},\quad \psi^{(1)}_{\mathcal{E}_{2}}\propto\frac{u_{1}^{(1)}}{W(u_{1}^{(1)},u_{2}^{(1)})}.
\end{equation}
 Due to a divergent behavior of the Wronskian when $|x| \rightarrow \infty$ which is stronger than the corresponding behavior of $u_{1}^{(1)}$, $u_{2}^{(1)}$,  the Hamiltonian $H^{(1)}$ has two new bound states, $\psi^{(1)}_{\mathcal{E}_{1}}$ and $\psi^{(1)}_{\mathcal{E}_{2}}$, so its spectrum becomes Sp$\lbrace H^{(1)} \rbrace =\lbrace E_{-2}, \mathcal{E}_{1}, E_{n},~n=0,~1,~2,\dots  \rbrace$.


There is an equivalent way to construct the same potential up to an additive constant, through another second-order transformation using different seed solutions. The previous transformation inserted $\mathcal{E}_{1},~\mathcal{E}_{2}$ in the spectrum, thus  the spectra of $H^{(1)}$ and $H$ differ only in the position of the ground state and the first excited level of $H^{(1)}$. We can obtain basically the same spectrum by deleting the first excited level and adding a new level at the right place of the harmonic oscillator spectrum. We will employ now the seed solutions 
\begin{equation} \label{u gamma}
   u_{1}^{(2)}(x)=\psi_{1}(x),\quad u_{2}^{(2)}(x)=e^{-\frac{x^2}{2}}[H_{\lambda_{2}}(x)+\gamma H_{\lambda_{2}}(-x)]
\end{equation}
 with factorization energies $\widetilde{\mathcal{E}}_{1}=E_{1}$ and $\widetilde{\mathcal{E}}_{2}=\mathcal{E}_{1}+2$, respectively,  thus  $E_{0}<\widetilde{\mathcal{E}_{2}}<E_{2}$, and $\lambda_{2}=\lambda_1+2$. Moreover,  we will take the same value of $\gamma$ as in the previous SUSY transformation. This will lead again to a nodeless Wronskian  $W(u_{1}^{(2)},u_{2}^{(2)})$ which can be written as
\begin{equation}
    W(u_{1}^{(2)},u_{2}^{(2)})=e^{-x^2}{\mathcal{W}}(x),
\end{equation}
with $\mathcal{W}(x)=W[H_{1}(x),H_{\lambda_{2}}(x)+\gamma H_{\lambda_{2}}(-x)]$. From equation \eqref{susypoten}, the SUSY partner potential can be expressed as
\begin{align}\label{wronsmath}
    V^{(2)}=\frac{1}{2}x^{2}+2-[\ln\mathcal{W}(x)]''.
\end{align}
Again, through relations \eqref{B+} and \eqref{B-}, we can define the second-order differential operators $B^{(2)},~B^{(2)+}$ which intertwine the Hamiltonian $H^{(2)}$ with $H$ as 
\begin{equation}
    H^{(2)}B^{(2)+}=B^{(2)+}H, \quad \quad H B^{(2)}=B^{(2)}H^{(2)}.
\end{equation}
The eigenfunctions of this Hamiltonian are
\begin{equation}
    \psi_{n}^{(2)}=\frac{B^{(2)+}\psi_{n}}{\sqrt{(E_{n}-\widetilde{\mathcal{E}}_{1})(E_{n}-\widetilde{\mathcal{E}}_{2})}},\quad n=0,2,\dots,
\end{equation}
and the missing states
\begin{equation}
    \psi^{(2)}_{\widetilde{\mathcal{E}}_{1}}\propto\frac{u_{2}^{(2)}}{W(u_{1}^{(2)},u_{2}^{(2)})},\quad \psi^{(2)}_{\widetilde{\mathcal{E}}_{2}}\propto\frac{u_{1}^{(2)}}{W(u_{1}^{(2)},u_{2}^{(2)})}.
\end{equation}
In this case, due to the divergent behavior of the solution $u_{2}^{(2)}$ when $|x|\rightarrow \infty$, the missing state $\psi_{\widetilde{\mathcal{E}}_{1}}^{(2)}$ is not square-integrable, but since $u_{1}^{(2)}$ converges, the state $\psi_{\widetilde{\mathcal{E}}_{2}}^{(2)}$ is normalizable. Therefore, the Hamiltonian $H^{(2)}$ has the discrete spectrum Sp$\lbrace H^{(2)} \rbrace = \lbrace E_{0},\widetilde{\mathcal{E}}_{2},E_{n},~n=2,~ 3,~ 4, \dots \rbrace$. We can see this transformation as we move the first excited state to an arbitrary place between the ground state and the second excited state \cite{david2010supersymmetric}.

\begin{figure}[t]
\centering
\includegraphics[scale=0.68, width=0.49\textwidth]{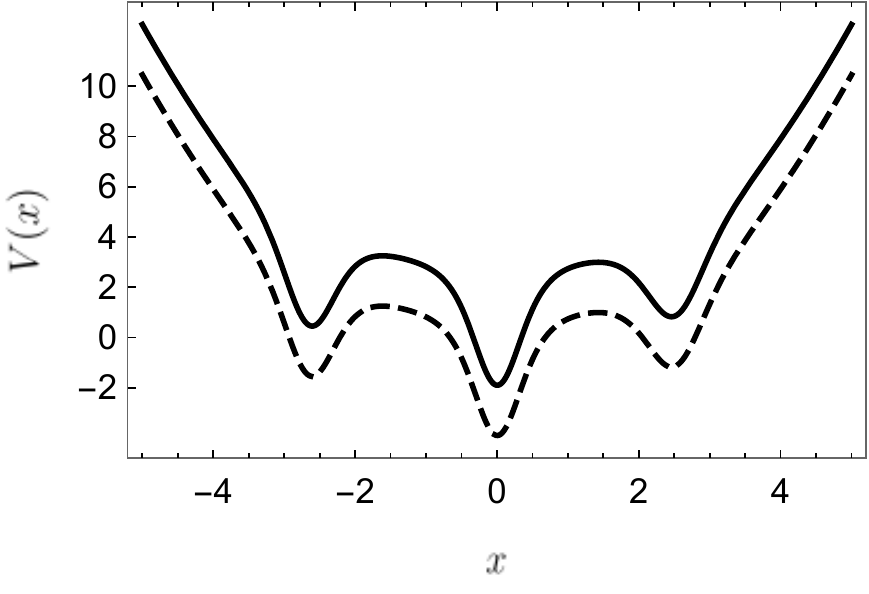}
\includegraphics[scale=0.68, width=0.49\textwidth]{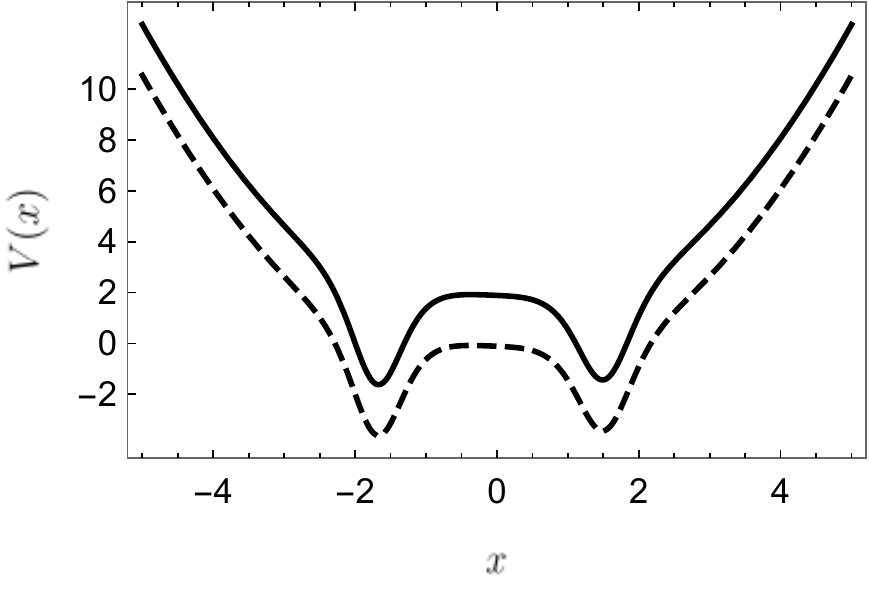}
\caption{Equivalent non-rational extensions of the harmonic oscillator for $\gamma=4$. \textbf{Left}: Potential $V^{(2)}$  (black continuous line) where the first excited state is moved down to $\widetilde{\mathcal{E}}_{2}=\frac{5}{9}$, and potential $V^{(1)}$ (dashed line) where two new levels are generated at $\mathcal{E}_{1}=-\frac{13}{9}$, and $\mathcal{E}_{2}=-\frac{3}{2}$. \textbf{Right}: Potential $V^{(2)}$ (black continuous line) where the first excited state is moved up to  $\widetilde{\mathcal{E}}_{2}=\frac{49}{20}$, and potential $V^{(1)}$ (dashed line) where two new levels are added at $\mathcal{E}_{1}=\frac{9}{20}$ and $\mathcal{E}_{2}=-\frac{3}{2}$.}
\label{field}
\end{figure}

It is important to notice that the seed solutions $u_1^{(1)},~ u_2^{(1)}$ used to construct $H^{(1)}$ are related to the seed solutions $u_1^{(2)},~ u_2^{(2)}$ giving place to $H^{(2)}$. The functions $u_2^{(1)}$ and $u_1^{(2)}$ satisfy $u_2^{(1)}= \sqrt{2 \sqrt{\pi}}~ e^{x^2} u_1^{(2)}$. Moreover,  $a^- a^- u_2^{(2)}=2\lambda_2 (\lambda_2-1) u_1^{(1)}$. These relations allow us to prove that $ \widetilde{V}^{(2)}=\widetilde{V}^{(1)}+2,$ (see Appendix), i.e.,
\begin{equation} \label{to prove 1}
    H^{(2)}=H^{(1)}+2.
\end{equation}
In Fig. (\ref{field}), we plot both SUSY generated potentials using these equivalent transformations. We observe they are the same, up to a constant equal to two.

In order to simplify the notation from now on, we will refer to the Hamiltonian $H^{(1)}$ simply as $\widetilde{H}$, $\mathcal{E}_{1} \rightarrow \epsilon$ and its eigenstates as $\lbrace \widetilde{\psi}_{E_{-2}},\widetilde{\psi}_{\epsilon},\widetilde{\psi}_{n},n=0,1,2,\dots \rbrace$. Then $-3/2 < \epsilon < 1/2$ is the position of the first excited energy level of $\widetilde{H}$, and $\gamma > 0$ is a parameter characterizing $\widetilde{H}$.  

Let us define now the fourth-order differential operators:
\begin{equation} \label{ladno}
\mathcal{L}^{+}=B^{(1)+}B^{(2)},\quad \mathcal{L}^{-}=B^{(2)+}B^{(1)}.
\end{equation}    
Since these operators fulfill the following commutation relations 
\begin{align}
    [\widetilde{H},\mathcal{L}^{\pm}]&=\pm 2 \mathcal{L}^{\pm},\\
    \nonumber [\mathcal{L}^{-},\mathcal{L}^{+}]&=(\widetilde{H}+2-\mathcal{E}_{1})(\widetilde{H}+2-\mathcal{E}_{2})(\widetilde{H}+2-\widetilde{\mathcal{E}}_{1})(\widetilde{H}+2-\widetilde{\mathcal{E}}_{2})\\
    &\quad -(\widetilde{H}-\mathcal{E}_{1})(\widetilde{H}-\mathcal{E}_{2})(\widetilde{H}-\widetilde{\mathcal{E}}_{1})(\widetilde{H}-\widetilde{\mathcal{E}}_{2}),
 \end{align}
they are in fact ladder operators that connect eigenstates whose energy levels differ by two energy units of $\widetilde{H}$. 
Moreover, a comparison of the previous equations with \eqref{PHA} leads us to conclude that the operators $\widetilde{H},~ \mathcal{L}^{-}, ~\mathcal{L}^{+} $ realize a polynomial Heisenberg algebra of third degree with $\omega=2$ and
\begin{equation}
N_{4}(\widetilde{H})=\mathcal{L}^{+}\mathcal{L}^{-}=(\widetilde{H}-\mathcal{E}_{1})(\widetilde{H}-\mathcal{E}_{2})(\widetilde{H}-\widetilde{\mathcal{E}}_{1})(\widetilde{H}-\widetilde{\mathcal{E}}_{2}).
\end{equation}
Finally, the kernel of the annihilation operator $\mathcal{L}^{-}$ is generated by the functions
\begin{equation}
    K_{\mathcal{L}^{-}}=\lbrace \widetilde{\psi}_{E_{-2}},\widetilde{\psi}_{\epsilon}, \widetilde{\psi}_{1},B^{(1)+}u_{2}^{(2)} \rbrace. 
\end{equation}

\begin{figure}[t]
\centering
\includegraphics[scale=0.5]{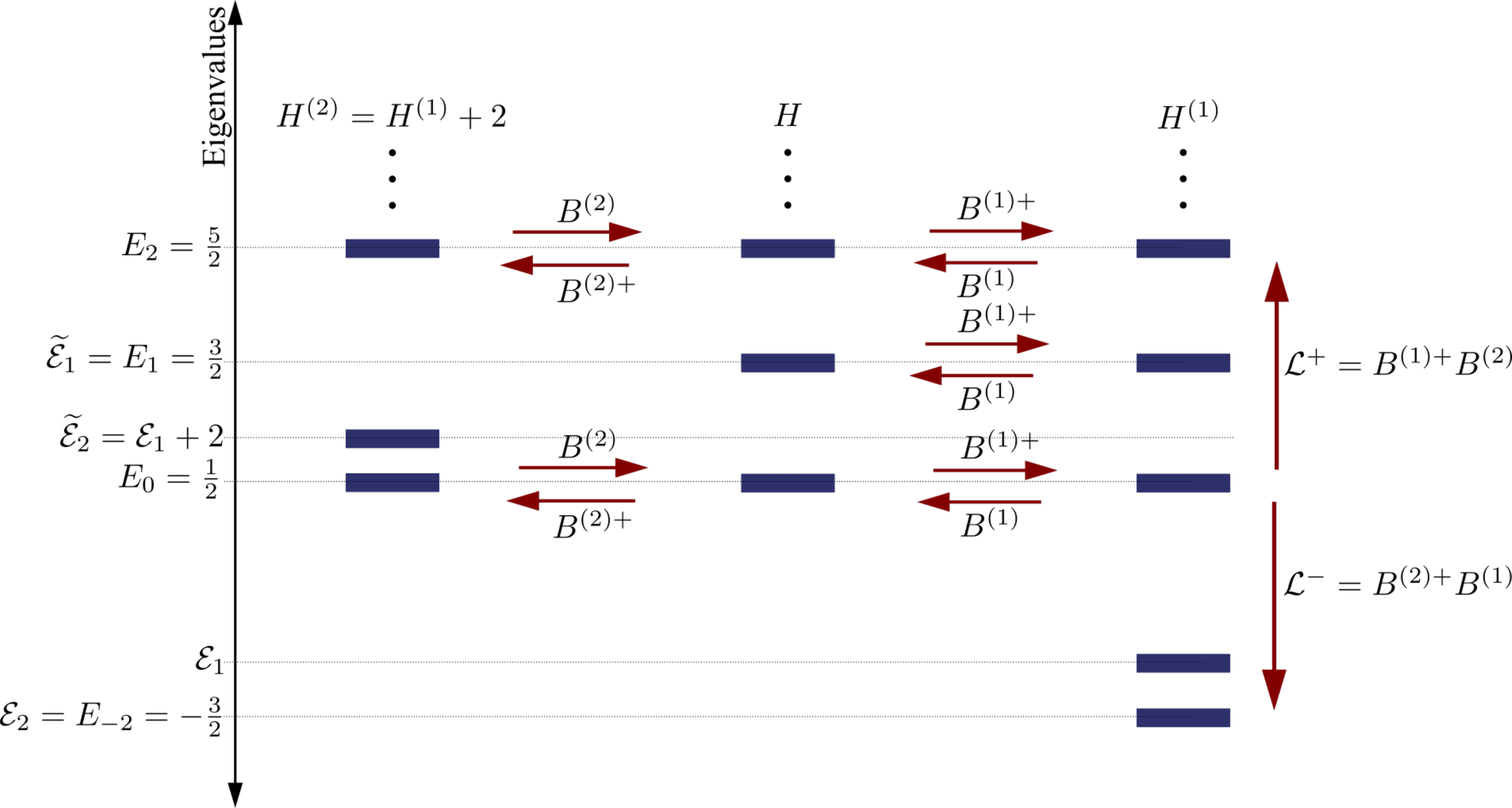}
\caption{Diagram of the action of the two-step ladder operators $\mathcal{L}^{-},\mathcal{L}^{+}$ defined in Eq. (\ref{ladno}).}
\label{figuritanon}
\end{figure}

In Fig. \ref{figuritanon} we can see a diagram of the action of the two-step ladder operators $\mathcal{L}^{-},\mathcal{L}^{+}$ and the intertwining operators $B^{(2)+},~B^{(2)},~B^{(1)+},B^{(1)}$ onto the eigenstates of $H^{(2)}, ~H, ~H^{(1)}$. The intertwining operators connect eigenfunctions horizontally (with the same energy). The composed action $B^{(1)+} B^{(2)}=\mathcal{L}^{+}$ moves horizontally  left-to-right from the energy ladder of $H^{(2)}$ to $H^{(1)}$. Since $H^{(2)}=H^{(1)}+2$, this is equivalent to an upward displacement of two units in the energy ladder of $H^{(1)}$.  In the same way, the action of $B^{(2)+} B^{(1)}=\mathcal{L}^{-}$ can be seen as an effective downward movement of two units in the ladder of $H^{(1)}$.

\section{Coherent states and their properties}

Now that we have identified the algebraic structure of the ladder operators $\mathcal{L}^{-},\mathcal{L}^{+}$, let us construct the Barut-Girardello coherent states in the standard way,
\begin{equation}
    \mathcal{L}^{-}\ket{z}=z\ket{z}, 
\end{equation}
where $z\in \mathbb{C}$. Recall that  $[\widetilde{H},\mathcal{L}^{\pm}]=\pm 2 \mathcal{L}^{\pm}$, meaning that the application of $\mathcal{L}^\pm$ onto an eigenstate of $\widetilde{H}$ will increase/decrease the energy in two units. Thus the Hilbert space $\mathbb{H}$ can be decomposed as the direct sum of two infinite dimension subspaces whose basis vectors are obtained by the successive action of $\mathcal{L}^+$ onto $\widetilde{\psi}_{E_{-2}}$ and $\widetilde{\psi}_1$ (labelled by the index $\nu=-2,~1$, i.e., $\mathbb{H}^{-2}$ and $\mathbb{H}^1$, respectively) plus the subspace $\mathbb{H}^{\epsilon}$ spanned by the single eigenstate $\widetilde{\psi}_{\epsilon}$. 

Since $\widetilde{\psi}_{\epsilon}$ belongs to the kernel $K_{\mathcal{L}^{-}}$, it constitutes by itself a Barut-Girardello coherent state with eigenvalue $z=0$. Let us expand now $\ket{z^{\nu}}$ in the basis of eigenstates of the corresponding subspace $\mathbb{H}^{\nu}$: 
\begin{equation}
    \ket{z^{\nu}}=\sum^{\infty}_{n=0}c_{n}\ket{\nu+2n},
\end{equation}
where we added the superindex $\nu$ to emphasize the subspace to which the coherent state belongs. By applying the operator $\mathcal{L}^{-}$ on both sides of the above equation, after some work it is found that the coefficients of the expansion are given by
\begin{equation}
c_{n}=\frac{(z/4)^{n}c_{0}}{\sqrt{\displaystyle\prod_{k=0}^{n}\left[k+\frac{2\nu-2\epsilon+1}{4}\right]\left[k+\frac{2\nu-2\epsilon-3}{4}\right]\left[k+\frac{\nu+2}{2}\right]\left[k+\frac{\nu-1}{2}\right]}}.\label{coefcohe}
\end{equation}
\normalsize
Thereupon, the coherent states can be written as                        
\begin{align}
\ket{z^{\nu}}= & c_0 \sum_{n=0}^{\infty}\left(\frac{z}{4}\right)^{n}  \sqrt{ \frac{\Gamma\left(\frac{2\nu-2\epsilon+5}{4}\right)     \Gamma\left(\frac{2\nu-2\epsilon+1}{4}\right)\Gamma\left(\frac{\nu+4}{2}\right)\Gamma\left(\frac{\nu+1}{2}\right)}{\Gamma\left(\frac{2\nu-2\epsilon+5}{4}+n\right)     \Gamma\left(\frac{2\nu-2\epsilon+1}{4}+n\right)\Gamma\left(\frac{\nu+4}{2}+n\right)\Gamma\left(\frac{\nu+1}{2}+n\right)} }\ket{\nu+2n}, \label{coherentnonex} 
\end{align}
where $c_{0}$ is the normalization constant given by
\begin{equation}
    c_{0}=\left[{}_{1}F_{4}\left(1;\frac{2\nu-2\epsilon+5}{4},\frac{2\nu-2\epsilon+1}{4},\frac{\nu+4}{2},\frac{\nu+1}{2};\frac{\abs{z}^{2}}{16}\right)\right]^{-1/2}, 
\end{equation}
with $_pF_q$ being a generalized hypergeometric series. We define now a function which will be useful later on
\begin{equation}\label{auxcnon}
   c_{0}(a,b)=\left[{}_{1}F_{4}\left(1;\frac{2\nu-2\epsilon+5}{4},\frac{2\nu-2\epsilon+1}{4},\frac{\nu+4}{2},\frac{\nu+1}{2};\frac{a^{*}b}{16}\right)\right]^{-1/2}.
\end{equation}
\normalsize
Let us analyze next some physical and mathematical properties of these coherent states.
\subsection{Completeness relation} 
    
We can show that the states \eqref{coherentnonex}   satisfy as well the fourth definition \eqref{fourth def} of coherent states, i.e., 
they fulfill a completeness relation of the form
\begin{equation} \label{id2}
\int_{\mathbb{C}}\ket{z^{\nu}}\bra{z^{\nu}}\mu(z)d^{2}z=\mathbb{1}_{\mathbb{H}^{\nu}}
\end{equation}
on each Hilbert subspace $\mathbb{H}^{\nu}$ spanned by the kets $\ket{\nu+2n}$, where $\mu(z)$ is a positive definite measure function. To find $\mu(z)$ we propose the ansatz 
\begin{equation}\label{med2}
\mu(z)=\frac{f(\abs{z})}{16 \pi c_{0}^{2}\Gamma\left(\frac{2\nu-2\epsilon+5}{4}\right)     \Gamma\left(\frac{2\nu-2\epsilon+1}{4}\right)\Gamma\left(\frac{\nu+4}{2}\right)\Gamma\left(\frac{\nu+1}{2}\right)},
\end{equation}
substitute it in equation (\ref{id2}), change to polar coordinates $z=re^{i\theta}$ and make the integral in $\theta$. At the end we arrive at the following moment problem
 \begin{align}
     \nonumber   \int^{\infty}_{0}(r^{2}/16)^{n}f(r)2rdr&=16 \Gamma\left(\frac{2\nu-2\epsilon+5}{4}+n\right)     \Gamma\left(\frac{2\nu-2\epsilon+1}{4}+n\right)\\
     &\quad \times\Gamma\left(\frac{\nu+4}{2}+n\right)\Gamma\left(\frac{\nu+1}{2}+n\right).
\end{align}
Finally, by changing the variable $y=r^{2}$ and the summation index $n=s-1$, we arrive at
    \begin{align}
    \nonumber \int_{0}^{\infty}(y/16)^{s-1}f(y)dy/16&= \Gamma\left(\frac{2\nu-2\epsilon+1}{4}+s\right)     \Gamma\left(\frac{2\nu-2\epsilon-3}{4}+s\right)\\
     &\quad \times\Gamma\left(\frac{\nu+2}{2}+s\right)\Gamma\left(\frac{\nu-1}{2}+s\right).\label{fnon}    
    \end{align}
  It follows that $f(y)$ is the inverse Mellin transform of the right hand side of the equation (\ref{fnon}), which can be written in terms of the Meijer-G function as follows:
     \begin{equation}\label{Meijerfunc}
       \MeijerG{\alpha,\beta}{\kappa,\iota}{ a_{1},...,a_{\kappa} \\ b_{1},...,b_{\iota} } { x } \equiv \mathcal{M}^{-1}\left[\frac{\displaystyle\prod_{j=1}^{\alpha}\Gamma(b_{j}+s)\displaystyle\prod_{j=1}^{\beta}\Gamma(1-a_{j}-s)}{\displaystyle\prod_{j=\alpha+1}^{\iota}\Gamma(1-b_{j}-s)\displaystyle\prod_{j=\beta+1}^{\kappa}\Gamma(a_{j}+s)};x\right].
    \end{equation}
    In our case, $x \to y/16$, $\alpha=4$, $\beta=0$, $\kappa=0$, $\iota=4$, $b_{1}=\frac{2\nu-2\epsilon+1}{4}$, $b_{2}=\frac{2\nu-2\epsilon-3}{4}$, $b_{3}=\frac{\nu+2}{2}$, and $b_{4}=\frac{\nu-1}{2}$, i.e.
   \begin{equation} \label{f1non} 
      f(y)=\MeijerG{4,0}{0,4}{ \frac{2\nu-2\epsilon+1}{4}, \frac{2\nu-2\epsilon-3}{4},\frac{\nu+2}{2},\frac{\nu-1}{2} } { \frac{y}{16} }.  
    \end{equation}
Thus, by inserting this function in equation (\ref{med2}), we obtain that our coherent states satisfy the required completeness relation. 

To prove the positiveness of the function $\mu(z)$, we employ the convolution property of the inverse Mellin transform, also called general Parseval formula, which is given by
\begin{equation}\label{parseformula}
        \mathcal{M}^{-1}[g^{*}(s)h^{*}(s);y]=\frac{1}{2\pi i}\int ^{+i\infty}_{-i\infty}g^{*}(s)h^{*}(s)x^{-s}dy=\int^{\infty}_{0}g(y t^{-1})h(t)t^{-1}dt.
    \end{equation}
For our case, we take
 \begin{equation}
    \nonumber g^{*}(s)=\Gamma\left(\frac{2\nu-2\epsilon+1}{4}+s\right)\Gamma\left(\frac{2\nu-2\epsilon-3}{4}+s\right), \quad h^{*}(s)=\Gamma \left(\frac{\nu+2}{2}+s\right)\Gamma\left(\frac{\nu-1}{2}+s\right),\\
    \end{equation}
    \normalsize
and use equation ($\ref{Meijerfunc}$), to obtain 
\begin{equation}\label{posmeijer}
    g(y)=\MeijerG{2,0}{0,2}{ \frac{2\nu-2\epsilon+1}{4}, \frac{2\nu-2\epsilon-3}{4}} {y}, \quad
    h(y)=\MeijerG{2,0}{0,2}{ \frac{\nu+2}{2}, \frac{\nu-1}{2}} {y}.
    \end{equation} 
Since we can express the Meijer-G functions of Eq. (\ref{posmeijer}) as \cite{erdelyi1953higher}
\begin{equation}\label{meijerbessel}
       \MeijerG{2,0}{0,2}{a,b} {x}=2x^{\frac{a+b}{2}}\mathcal{K}_{a-b}(2x^{1/2}),
\end{equation}
    where $\mathcal{K}_{\tau}(z)$ is the modified Bessel function of third kind represented by \cite{abramowitz}
\begin{align}\label{besselfunc}
     \mathcal{K}_{\tau}(z)\equiv\frac{\sqrt{\pi}}{\Gamma\left(\tau+1/2\right)}\left(\frac{z}{2}\right)^{\tau}\int^{\infty}_{1}e^{-zp}(p^{2}-1)^{\tau-\frac{1}{2}}dp,
    \quad \tau>-\frac{1}{2},\quad -\frac{\pi}{2}<\text{arg}~z<\frac{\pi}{2},
\end{align}
it turns out that the previous functions $g(y)$, and $h(y)$ can be written as
\begin{align}
    \label{gparse}g(y)&=4y^{\frac{2\nu-2\epsilon+3}{4}}\int^{\infty}_{1}e^{-2y^{1/2}p}(p^{2}-1)^{1/2}dp,\\
    \label{hparse}h(y)&=2\sqrt{\pi} y^{\frac{2\nu+7}{4}}\int^{\infty}_{1}e^{-2y^{1/2}p}(p^{2}-1)dp.
\end{align}
It can be seen now that $g(y)$, $h(y)\geq 0$ for $x\in[0,\infty)$. Then, by inserting (\ref{gparse}) and (\ref{hparse}) in equation (\ref{parseformula}), the positiveness of $f(y)$ in equation (\ref{f1non}) is guaranteed, and consequently $\mu(z)$ is a positive definite function.

 \subsection{Continuity on the label} 

\begin{figure}[t]
\centering
\includegraphics[scale=0.8, width=0.47\textwidth]{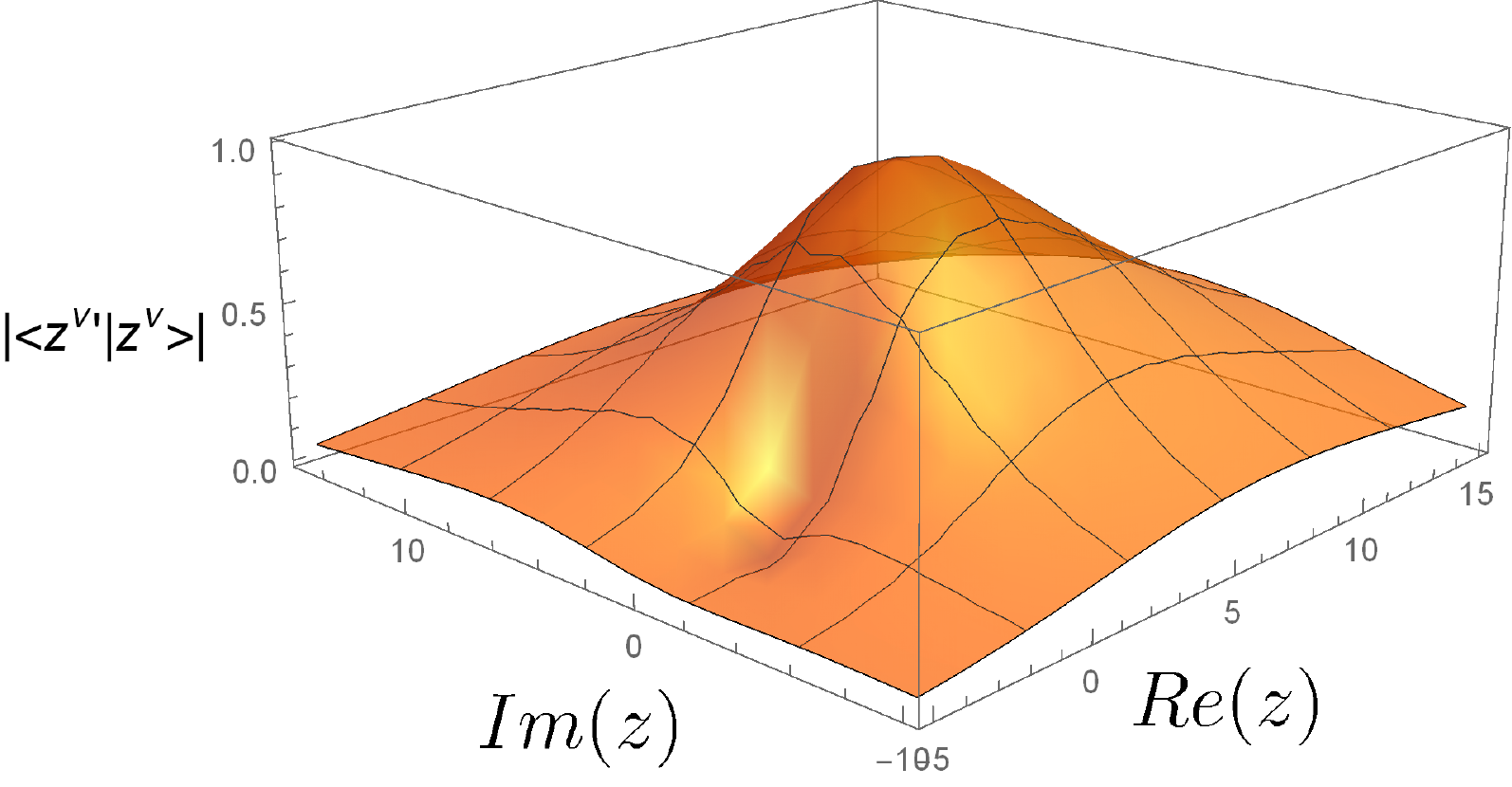}
\includegraphics[scale=0.8, width=0.47\textwidth]{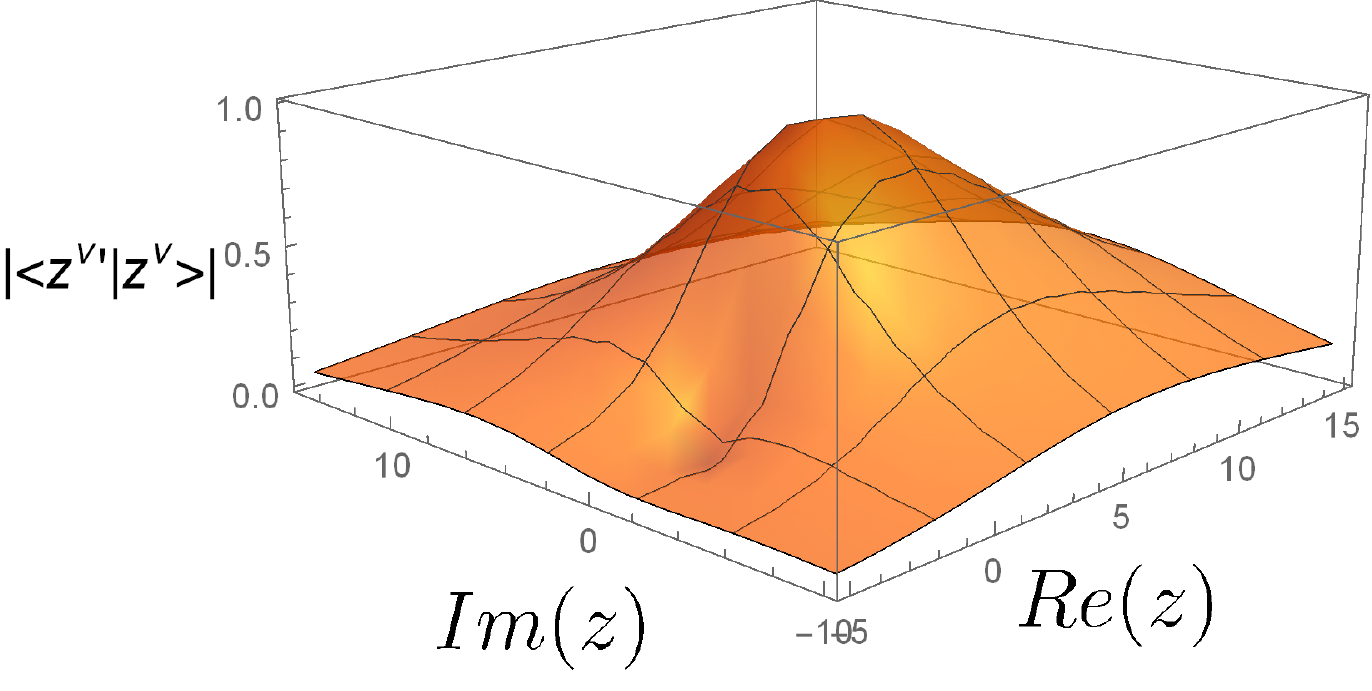}
\caption{Modulus of the scalar product $\bra{z^{'\nu}}\ket{z^{\nu}}$ of the coherent states $\ket{z^{\nu}}$ and $\ket{z^{'\nu}}
$ for $z'=4+i$ in the subspace with $\nu=-2$ (\textbf{left}), and $\nu=1$ (\textbf{right}).}
\label{projection}
\end{figure}

 
 By writing down the scalar product of two coherent states of the same subspace in terms of the auxiliary function (\ref{auxcnon}) 
    \begin{equation}
        \bra{z^{'\nu}}\ket{z^{\nu}}=\frac{c_{0}(z',z')c_{0}(z,z)}{c_{0}^{2}(z',z)},
    \end{equation}
we realize that if $z'\to z$ then $\ket{z^{'\nu}}\to\ket{z^{\nu}}$, and the absolute value of such scalar product goes to one, as we show in Fig. \ref{projection}. For the harmonic oscillator standard coherent states the behavior of the corresponding modulus of the projection is a Gaussian function \cite{glauber1963coherent}. As we can see, we have gotten somehow a similar behavior.

  \subsection{Mean energy values} 
\begin{figure}[t] 
\centering
\includegraphics[scale=0.8, width=0.47\textwidth]{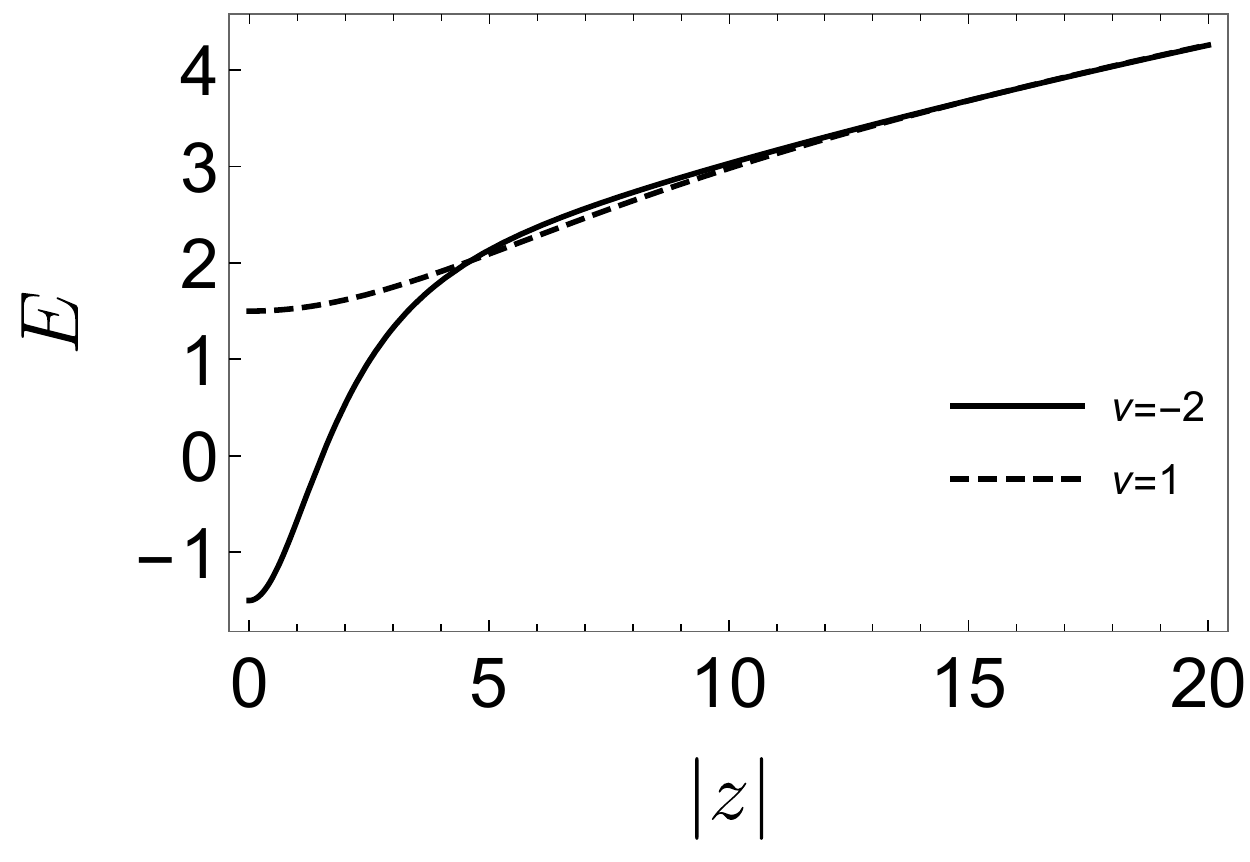}
\includegraphics[scale=0.8, width=0.47\textwidth]{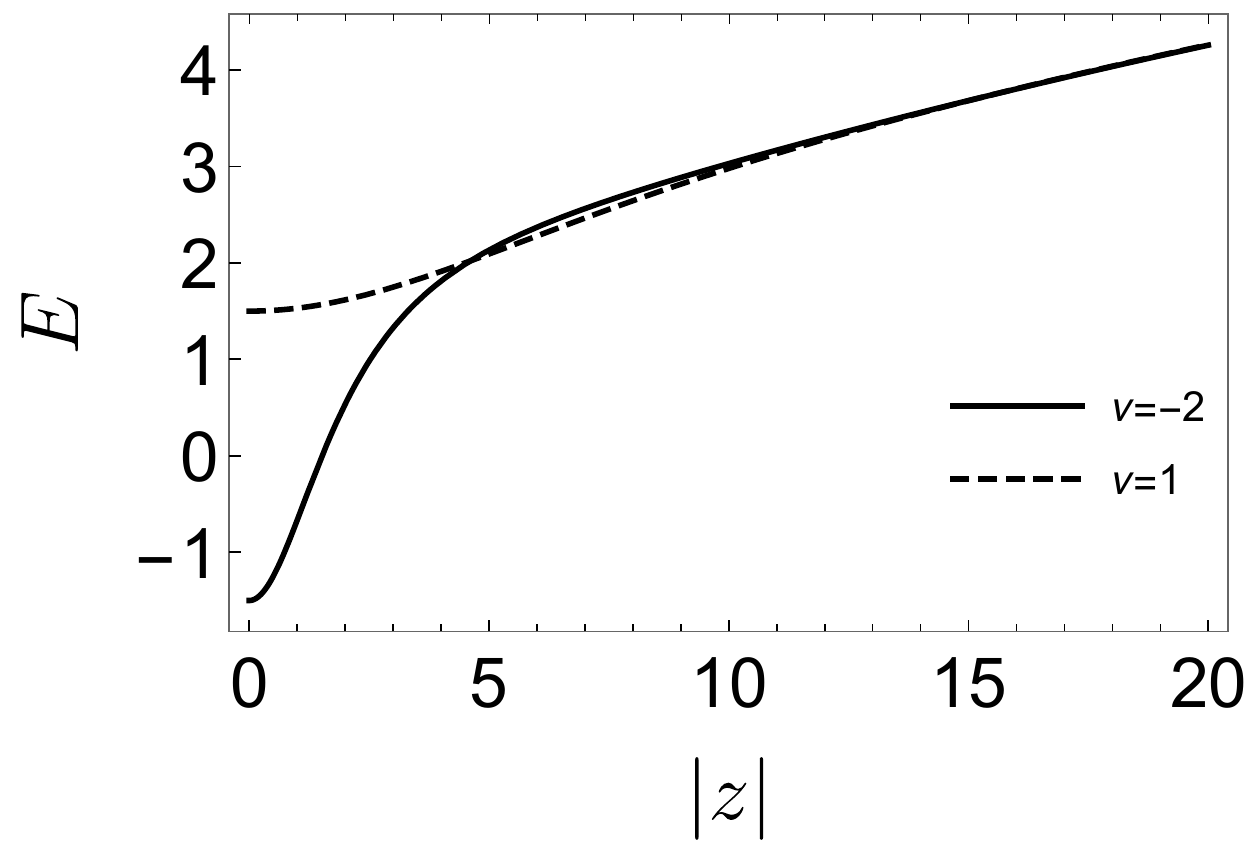}
\caption{Mean energy values as function of $\abs{z}$ in both subspaces for $\epsilon=-\frac{1}{4}$ (\textbf{left}), and $\epsilon=-\frac{3}{4}$ (\textbf{right}).}
\label{caso2ev}
\end{figure}

  The mean energy value is given by
  \begin{align}
  \nonumber \bra{z^{\nu}}\widetilde{H}\ket{z^{\nu}}&=\nu+\frac{1}{2}+\frac{8\abs{z}^{2}}{(2\nu-2\epsilon+5)(2\nu-2\epsilon+1)(\nu+1)(\nu+4)}\\
  &\quad\times\frac{ {}_{1}F_{4}\left(2; \frac{2\nu-2\epsilon+8}{4}, \frac{2\nu-2\epsilon+5}{4},\frac{\nu+6}{2},\frac{\nu+3}{2}; \frac{\abs{z}^{2}}{16}      \right)  }{ {}_{1}F_{4}\left(1; \frac{2\nu-2\epsilon+5}{4}, \frac{2\nu-2\epsilon+1}{4},\frac{\nu+5}{2},\frac{\nu+1}{2}; \frac{\abs{z}^{2}}{16}      \right)   }.
  \end{align}
We have plotted in Fig. \ref{caso2ev} such mean energy value as function of $\abs{z}$. We observe that the growth is slower than for the harmonic oscillator standard coherent states, which is quadratic in $|z|$ \cite{glauber1963coherent}. Moreover, it can be seen that for $|z|$ small, $\bra{z^{\nu}}\widetilde{H}\ket{z^{\nu}} \rightarrow \nu + 1/2$. 

\subsection{Temporal stability} 
By applying the time evolution operator, $U(t)=e^{-i\widetilde{H}t}$, onto the states (\ref{coherentnonex}), it is found the corresponding temporal evolution

\begin{align}
    U(t)\ket{z^{\nu}}&=e^{-i\left(\nu+\frac{1}{2}\right)t}\sum^{\infty}_{n=0}\frac{(ze^{-2it}/4)^{n}c_{0}}{\sqrt{\left(\frac{2\nu-2\epsilon+5}{4}\right)_{n}\left(\frac{2\nu-2\epsilon+1}{4}\right)_{n}\left(\frac{\nu+4}{2}\right)_{n}\left(\frac{\nu+1}{2}\right)_{n}}}\ket{\nu+2n}=e^{-iE_{0,\nu}t}\ket{\left(ze^{-i2t}\right)^\nu },
\end{align}
where $E_{0,\nu}=\nu+1/2$. Up to global phase factor, they evolve in time into another coherent states in the same subspace. As a first indication of their non-classical behavior, it can be seen that the period of every cyclic evolution is $\tau=\pi$, half the period of the standard coherent states.

\subsection{Time evolution of probability densities}

\begin{figure}[t] 
\centering
\includegraphics[scale=0.9, width=0.25\textwidth]{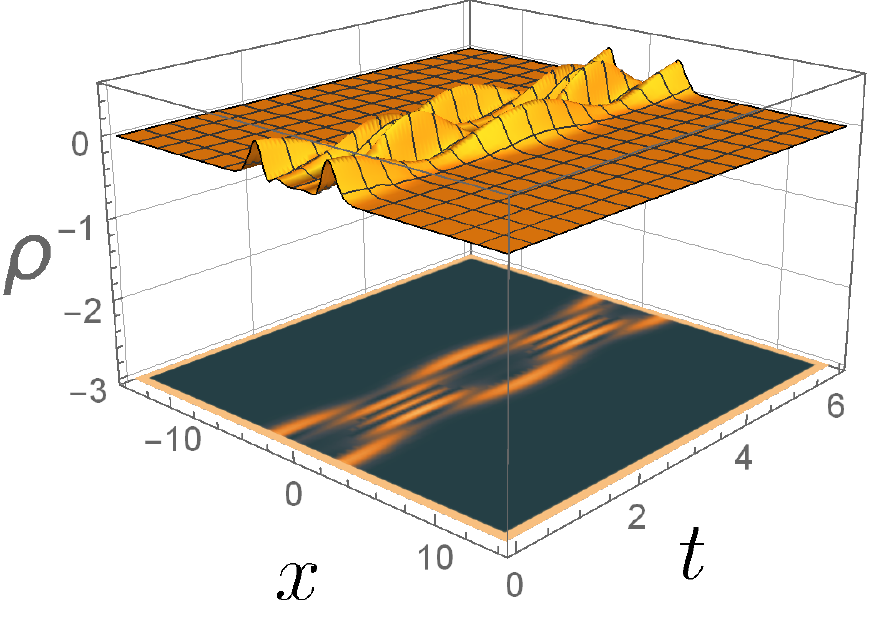}
\includegraphics[scale=0.9, width=0.25\textwidth]{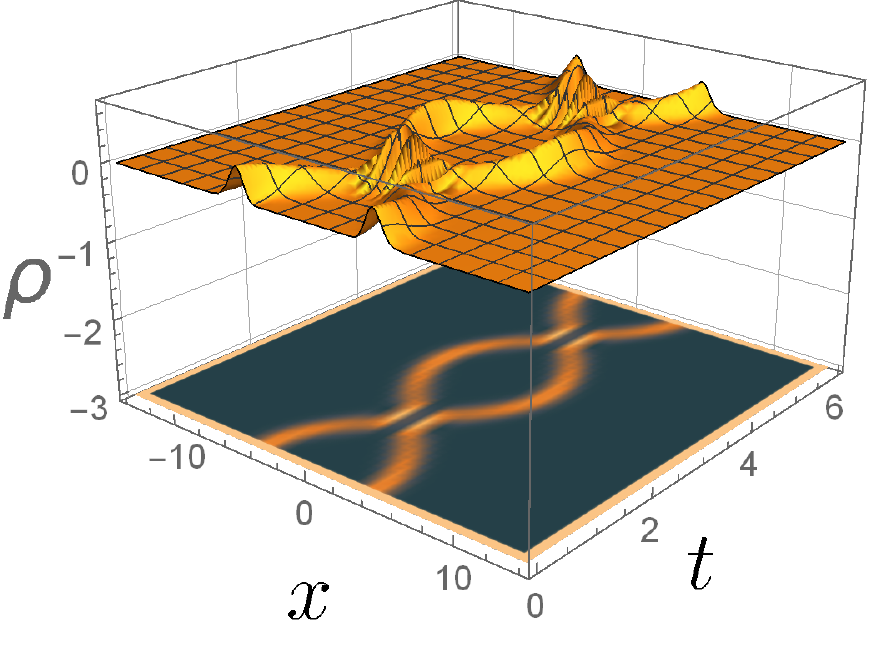}
\includegraphics[scale=0.9, width=0.25\textwidth]{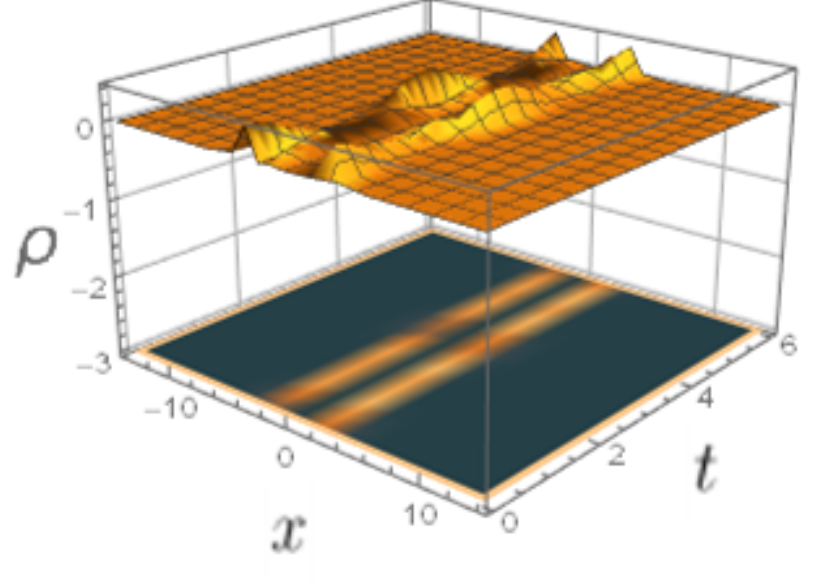}
\includegraphics[scale=0.9, width=0.23\textwidth]{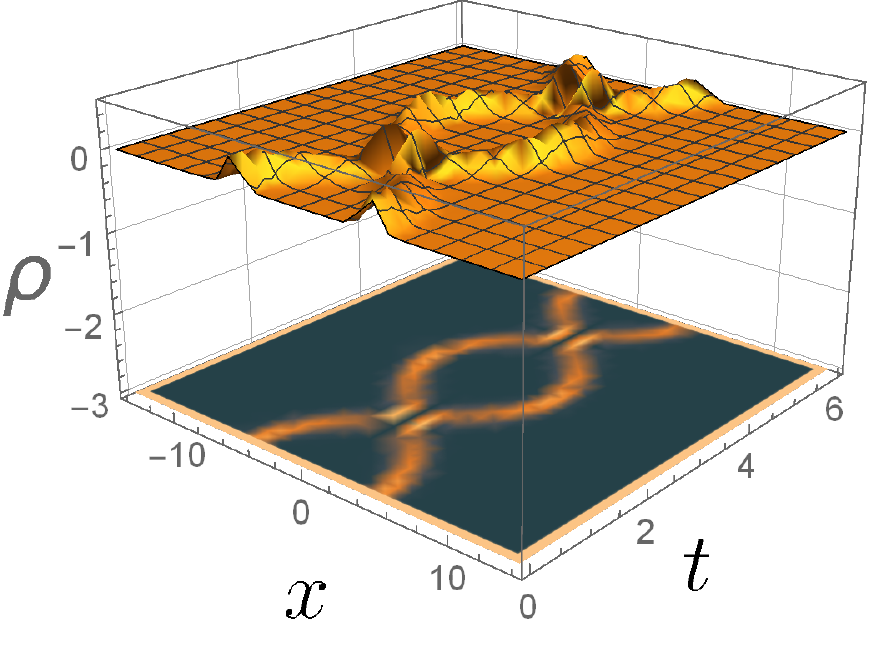}
\caption{Probability densities of the coherent states with $\epsilon=0$, $\gamma=2$. \textbf{First:} $\nu=-2$, $z=10$; \textbf{Second:} $\nu=-2$, $z=100$; \textbf{Third:} $\nu=1$, $z=10$; \textbf{Fourth}: $\nu=1$, $z=100$.} 
\label{plotdenposi}
\end{figure}

Let us analyze the time evolution of the coherent states probability density. We know that for the standard coherent states of the harmonic oscillator, this quantity is an oscillating Gaussian wave packet. On the other hand, for our coherent states, we obtain: 
\begin{align}\label{densitiesnon}
   \rho_z (x,t)=\abs{\bra{x}U(t)\ket{z^{\nu}}}^{2}= c_0^2 \left|\sum^{\infty}_{n=0}\frac{(ze^{-2it}/4)^{n}}{ \sqrt{\left(\frac{2\nu-2\epsilon+5}{4}\right)_{n} \left(\frac{2\nu-2\epsilon+1}{4}\right)_{n} \left(\frac{\nu+2}{2}+4\right)_{n} \left(\frac{\nu+1}{2}\right)_{n}   }      } \widetilde{\psi}_{\nu+2n}(x) \right|^{2},
\end{align}
where $\widetilde{\psi}_{\nu+2n}(x)=\bra{x}\ket{\nu+2n}$. This series falls off rapidly as $n$ and $|z|$ grow. We show plots of the probability densities (\ref{densitiesnon}) in Fig. \ref{plotdenposi}, where we cut-off the series at $n=12$ and take the values $z=10$ and $z=100$. We observe that for small values of $\abs{z}$ we do not get a clear pattern, but as we increase this quantity, each coherent state becomes composed by two wavepackets with a back-and-forth motion resembling a semi-classical behavior, since each individual wavepacket looks like a harmonic-oscillator coherent state. The two wavepackets interfere with each other, which is more notorious when they collide around $x=0$. A parity symmetry $x\rightarrow-x$ is only apparent, and it cannot be guaranteed for any SUSY extension, since the potential $\widetilde{V}$ is only symmetric around $x=0$ when $\gamma=0$.

\subsection{Wigner distributions}
\begin{figure}[t]
\centering
\includegraphics[scale=0.8,width=0.31\textwidth]{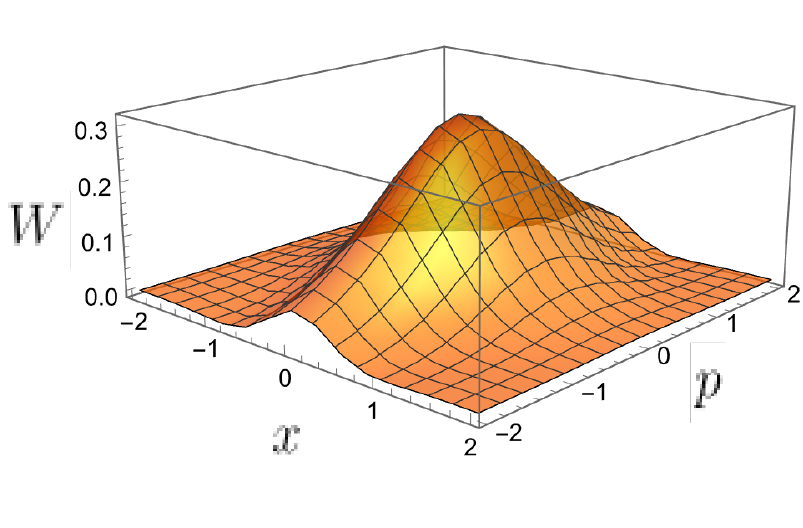}
\includegraphics[scale=0.8,width=0.31\textwidth]{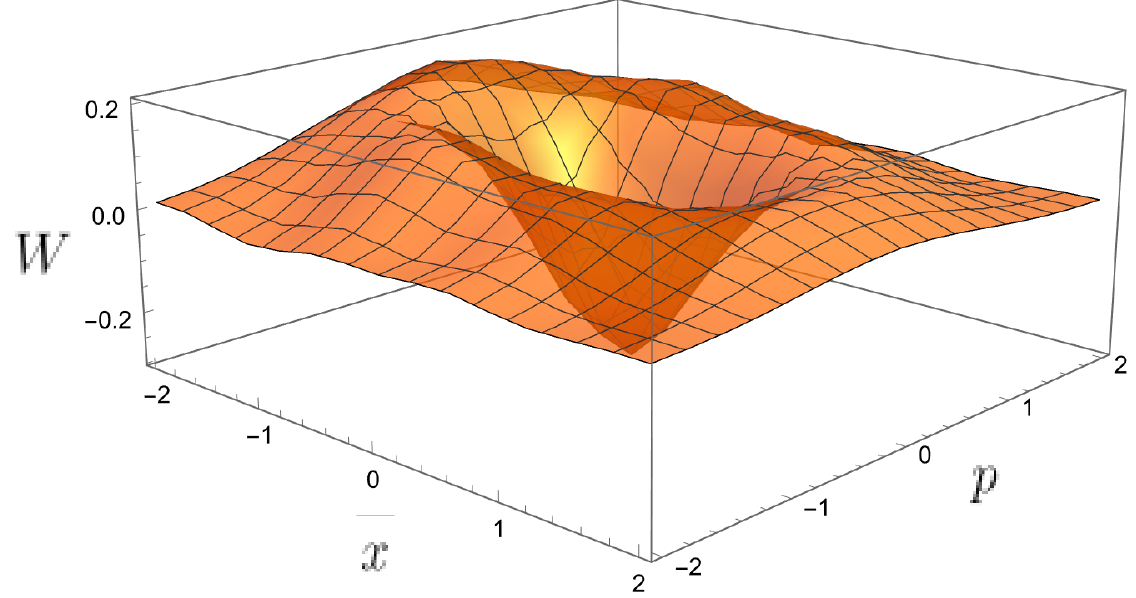}
\includegraphics[scale=0.8,width=0.31\textwidth]{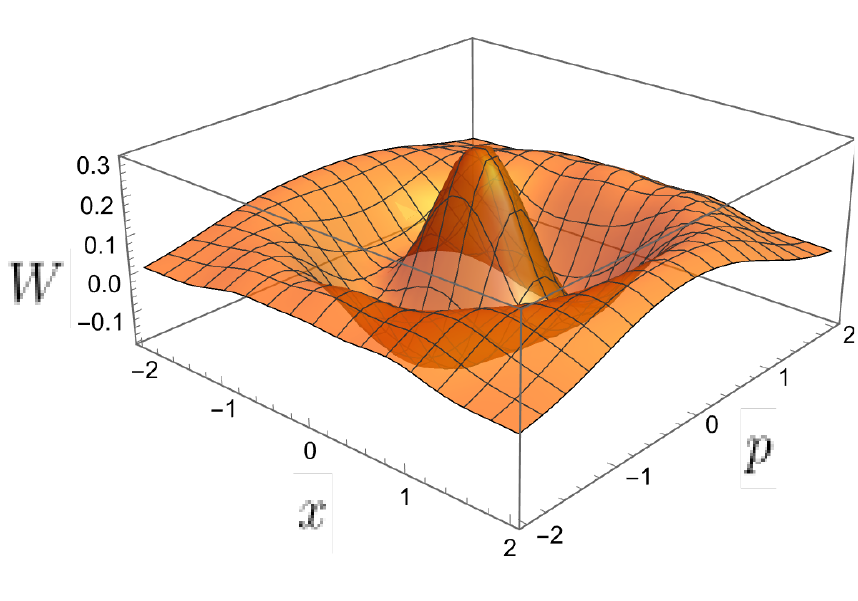} \\ 
\includegraphics[scale=0.8, width=0.40\textwidth]{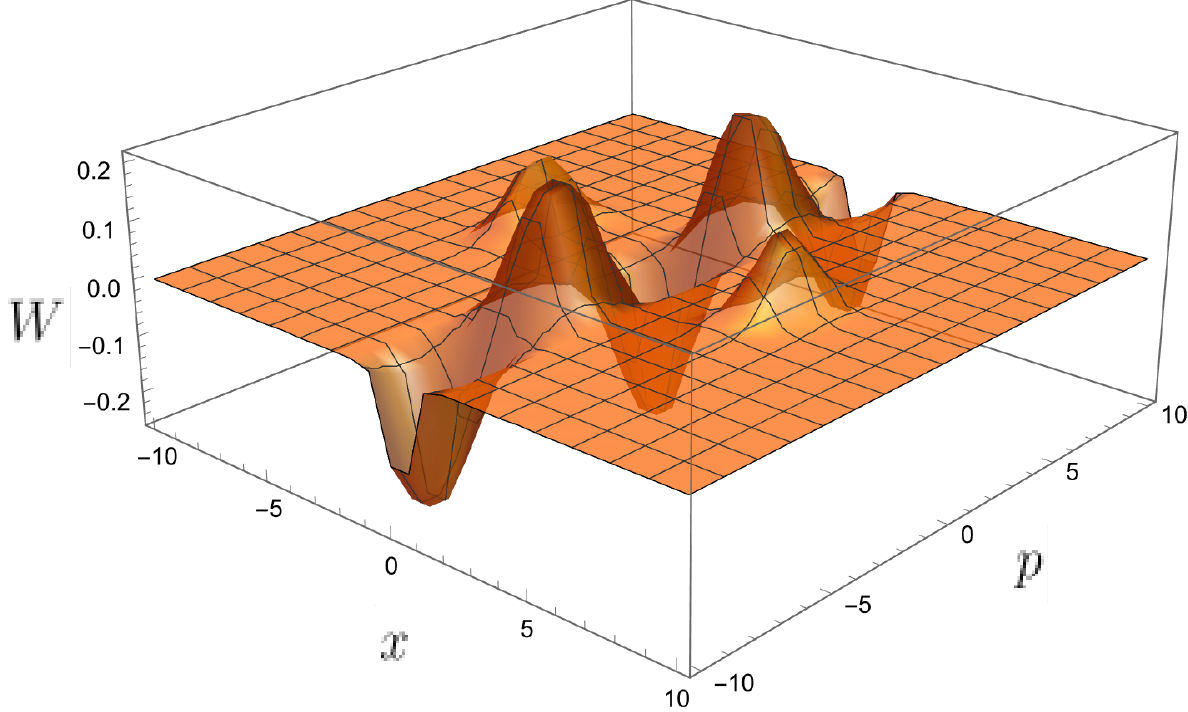}
\includegraphics[scale=0.9, width=0.40\textwidth]{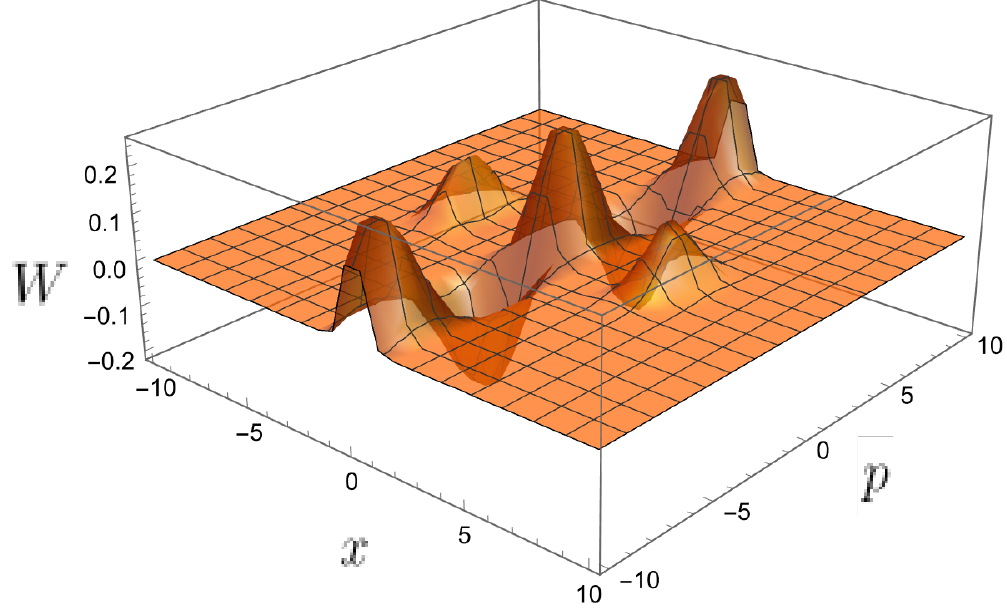}
\caption{Wigner functions of the ground state $\widetilde{\psi}_{E_{-2}}$ (\textbf{top left}), the first excited state $\widetilde{\psi}_{\epsilon}$ (\textbf{top center}), the second  excited state $\widetilde{\psi}_{0}$ (\textbf{top left}), coherent state with $\nu=-2$, $z=100$ (\textbf{bottom left}), and the coherent state with $\nu=1$, $z=100$ (\textbf{bottom right}).  The parameters employed are $\epsilon=0$, $\gamma=2$. }
\label{wigfocknon}
\end{figure}

 A valuable tool to study the nature of quantum states and to supply their phase space description is the Wigner quasiprobability distribution, which is given by  \cite{PhysRev.40.749}
\begin{equation}\label{wignerd}
        W(x,p)\equiv \frac{1}{2\pi}\int_{-\infty}^{\infty}\psi^{*}\left(x-\frac{y}{2}\right)\psi\left(x+\frac{y}{2}\right)e^{ipy}dy.
    \end{equation}
If the Wigner function of a quantum state has only positive values, it may be thought of as a quasiclassical state and can conform classical expectations.
In Fig. (\ref{wigfocknon}) top row, we plot the Wigner function of the first three eigenstates of the supersymmetric non-rational extended Hamiltonian $\widetilde{H}$. They behave similarly as the corresponding Wigner functions of the first three Fock states. On the other hand, the bottom row shows the Wigner functions for two coherent states, where we cut off the series at $n=12$ to compute numerically such distributions. We see that they have regions where non-positive values arise, which indicate a non-classical behavior of these states. 
  
\subsection{Mandel Q parameter}
Finally, let us investigate the statistics of these states through the Mandel $Q$ parameter, defined by \cite{mandel1979sub}
\begin{equation}
        Q=\frac{\langle N^{2} \rangle-\langle N\rangle^{2} }{\langle N \rangle}-1
    \end{equation}
 where $N$ is a number operator.
 
In general, depending on the value of $Q$, we have three different regimes for the statistical distributions: $Q>0$ means to have a super-Poissonian statistics such that $\bigtriangleup N^{2} >\langle N \rangle$ (the thermal light exhibits this statistics),  $Q=0$ a Poissonian distribution is obtained (as the classical coherent light) while $Q<0$  means to have a sub-Poissonian distribution such that $\bigtriangleup N ^{2}<\langle N \rangle$, which is related with a pure quantum behavior (an example being the squeezed light). By choosing now the number operator such that it retrieves the number excitation
     \begin{equation}
        N^{(\nu)}\ket{\nu+2n}= n\ket{\nu+2n},
    \end{equation}
the expectation values in the coherent states turn out to be
    \begin{equation}
    \langle N^{(\nu)} \rangle=\sum^{\infty}_{n=0}\abs{c_{n}}^{2}n,\quad \langle N^{(\nu) 2} \rangle=\sum^{\infty}_{n=0}\abs{c_{n}}^{2}n^{2},
    \end{equation}
where $c_{n}$ are the coefficients given in equation (\ref{coefcohe}).
    
\begin{figure}[t]
\centering
\includegraphics[scale=0.68, width=0.72\textwidth]{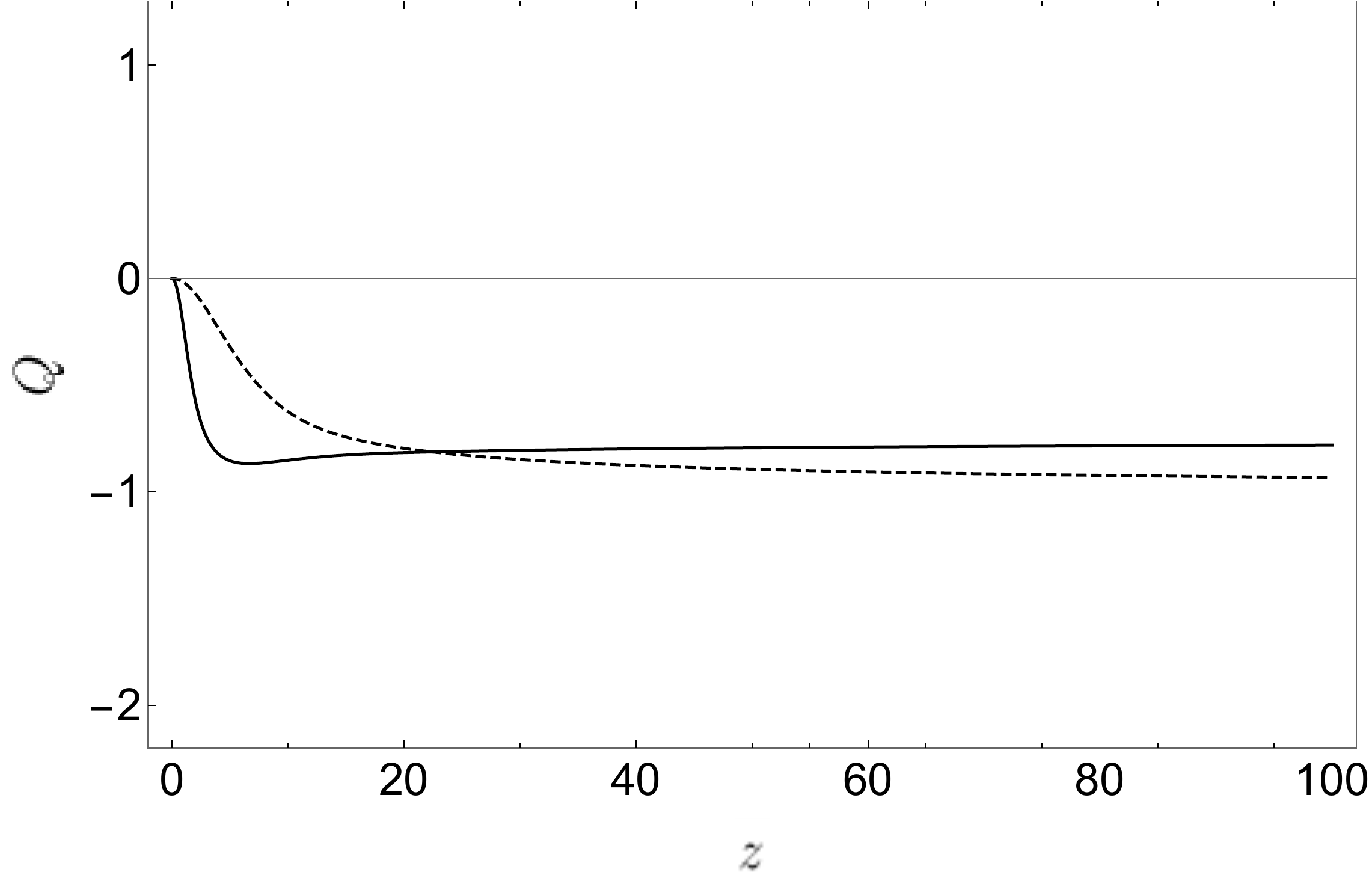}
\caption{Mandel Q parameter for the coherent states (\ref{coherentnonex}) with $\epsilon=0$, and $\gamma=2$; $\nu=-2$ (black line) and $\nu=1$ (dashed line).}
\label{mandelqn}
\end{figure}

In Fig. \ref{mandelqn} we have plotted the Mandel $Q$ parameter for our two sets of coherent states. We see that they exhibit a Poissonian statistics at $\abs{z}=0$ only, while a sub-Poissonian statistics for greater values of $\abs{z}$ is observed.

\section{Concluding remarks}

We have found a family of equivalent non-rational extensions of the harmonic oscillator, built through two different SUSY transformations. In one of them, two energy levels were created below the harmonic oscillator ground state; in the other transformation, the first excited state energy was moved to an arbitrary level between the ground state and the second excited state. For convenience, we expressed the general solution of the stationary Schr\"odinger equation in terms of Hermite functions. Both SUSY transformations generate the same family of potentials $\Tilde{V}$ with two free parameters: the factorization energy $\epsilon$ and the parameter $\gamma$. Then, we build fourth-order differential ladder operators for $\widetilde{H}$ as the product of the intertwining operators associated with both transformations. From these operators, we derived two sets of Barut-Girardello coherent states and studied their properties. We found that they are temporally stable cyclic states with a period equal to $\pi$, i.e. half the standard coherent states period, which represents the first sign of non-classical behavior. We showed as well that they satisfy a completeness relation. In addition, the associated mean-energy values increase slower than the corresponding result for the standard coherent states as $\abs{z}$ grows. For the evolution of the probability densities, we found that as the parameter $\abs{z}$ increases, a pattern-structure of two wave-packets oscillating and colliding at the center of the potential arises. This behavior is similar to the corresponding one for the coherent states of rational extensions \cite{hoffmann2018coherent}. Regarding Wigner functions, we observed that the distributions for the first three  eigenstates of the new Hamiltonian $\tilde{H}$ are qualitatively similar to the corresponding states of the oscillator. However, the Wigner functions for the coherent states possess regions with negative values. Furthermore, the Mandel $Q$ parameter showed sub-Poissonian statistics, which represents another sign of the quantum nature of these states.

\section*{Acknowledgments}
The authors acknowledge the support of Conacyt, grant FORDECYT-PRONACES/61533/2020.

\appendix
\section{Appendix}
To prove that $ H ^{(2)}= H ^{(1)}+2$ (see Eq. \eqref{to prove 1}), let us first note that $u_2^{(1)}$ and $u_1^{(2)}$ are related via the annihilation operator $a^-$ as follows: 
\begin{equation} \label{clave}
    u_{1}^{(1)}= \frac{1}{2 \lambda_2 (\lambda_2-1)} a^{-}a^{-}~u_{2}^{(2)} = \frac{1}{2 \lambda_2 (\lambda_2-1)} \left[ (x^{2} - \lambda_2) u_{2}^{(2)}+ x u_{2}^{(2)}{}'  \right],
\end{equation} 
where in the last equality we substituted the second derivative using Schr\"odinger equation, $u_2^{(2)}{}'' = (x^2 -\lambda_2 -1)u_2^{(2)}$. To simplify notation, let us define $W_1 = W(u_{1}^{(1)},u_{2}^{(1)})$ and $W_2 = W(u_{1}^{(2)},u_{2}^{(2)})$. Thus, we need to prove that \begin{eqnarray} \label{prove omega}
 H ^{(2)}- H ^{(1)}=\left(\ln W_2\right)'' - \left(\ln W_1\right)''= \left( \frac{W_2' W_1 - W_2 W_1'}{W_1 W_2} \right)'= \Omega'=2. 
\end{eqnarray}

The first step is to express $\Omega$ in terms of $u_i^{(j)};~i,j=1,2$. Since each $u_i^{(j)}$ satisfies a Schr\"odinger equation, we can substitute once again the second derivatives:
\begin{eqnarray}
\Omega=\frac{2 \lambda_2  u_{1}^{(2)} u_{1}^{(1)} u_{2}^{(1)} u_{2}^{(2)}{}'-2 u_{2}^{(2)} \left[ u_{1}^{(2)} \left(\lambda_2  u_{2}^{(2)} u_{1}^{(2)}{}'-(\lambda_2 -1) u_{1}^{(2)} u_{2}^{(1)}{}'\right)+(\lambda_2 -1) u_{1}^{(2)} u_{2}^{(1)} u_{1}^{(1)}{}'\right]}{\left(u_{2}^{(2)} u_{1}^{(2)}{}'-u_{1}^{(2)} u_{2}^{(2)}{}'\right) \left(u_{2}^{(1)} u_{1}^{(1)}{}'-u_{1}^{(1)} u_{2}^{(1)}{}'\right)}. 
\end{eqnarray}
Next, we use the explicit expressions of $u_1^{(2)}= \sqrt{2} ~\pi^{-1/4}~ x e^{-x^2/2}$ and $u_2^{(1)}=  2 x e^{x^2/2}$, to obtain: 
\begin{eqnarray}
\Omega=2 x \left(\frac{(\lambda_2 -1) u_2^{(2)}}{x u_2^{(2)}{}'+\left(x^2-1\right) u_2^{(2)}}+\frac{\lambda_2  u_1^{(1)}}{\left(x^2+1\right) u_1^{(1)}-x u_1^{(1)}{}'}\right).
\end{eqnarray}
Using expression \eqref{clave} in the last equation we further simplify to $\Omega= 2x$. Finally, substituting $\Omega=2x$ in equation \eqref{prove omega} we prove the relation $ H ^{(2)}= H ^{(1)}+2$.



\bibliographystyle{unsrt}
\bibliography{References}

\begin{thebibliography}{10}

\bibitem{matveev92}
V.B. Matveev{,} and M.A. Salle.
\newblock {\em Darboux Transformations and Solitons}.
\newblock Springer Series in Nonlinear Dynamics. Springer Berlin Heidelberg,
  1992.

\bibitem{Khare95}
F.~Cooper, A.~Khare, and U.~Sukhatme.
\newblock Supersymmetry and quantum mechanics.
\newblock {\em Physics Reports}, 251(5):267 -- 385, 1995.

\bibitem{Bagchi01}
B.~Bagchi.
\newblock {\em Supersymmetry in quantum and classical mechanics}.
\newblock Chapman \& Hall/CRC, Boca Raton, 2001.

\bibitem{Andrianov04}
A.~Andrianov{,} and F.~Cannata.
\newblock Nonlinear supersymmetry for spectral design in quantum mechanics.
\newblock {\em Journal of Physics A: Mathematical and General}, 37(43):10297,
  2004.

\bibitem{david2010supersymmetric}
D.J.~Fern{\'a}ndez C.
\newblock Supersymmetric quantum mechanics.
\newblock In {\em AIP Conference Proceedings}, volume 1287, pages 3--36.
  American Institute of Physics, 2010.

\bibitem{gangopadhyaya17}
A.~Gangopadhyaya, J.V. Mallow, and C.~Rasinariu.
\newblock {\em Supersymmetric Quantum Mechanics: An Introduction (Second
  Edition)}.
\newblock World Scientific Publishing Company, 2017.

\bibitem{junker19}
G.~Junker.
\newblock {\em Supersymmetric Methods in Quantum, Statistical and Solid State
  Physics}.
\newblock IOP Expanding Physics. Institute of Physics Publishing, 2019.

\bibitem{Kuru2009}
Ş. Kuru, J.~Negro, and L.M. Nieto.
\newblock {Exact analytic solutions for a Dirac electron moving in graphene
  under magnetic fields}.
\newblock {\em Journal of Physics: Condensed Matter}, 21(45):455305, 2009.

\bibitem{Midya2014}
Midya{,} B. and D.J. Fern{\'{a}}ndez~C.
\newblock {Dirac electron in graphene under supersymmetry generated magnetic
  fields}.
\newblock {\em Journal of Physics A: Mathematical and Theoretical},
  47(28):285302, 2014.

\bibitem{castillo2020dirac}
M.~Castillo-Celeita{,} and D.J.~Fernández C.
\newblock Dirac electron in graphene with magnetic fields arising from
  first-order intertwining operators.
\newblock {\em Journal of Physics A: Mathematical and Theoretical},
  53(3):035302, 2020.

\bibitem{contreras2020super}
A.~Contreras-Astorga{,}~F. Correa{,} and V~Jakubsk{\`y}.
\newblock Super-{K}lein tunneling of {D}irac fermions through electrostatic
  gratings in graphene.
\newblock {\em Physical Review B}, 102(11):115429, 2020.

\bibitem{daniel2020electron}
D.J. Fernández C.{,}~D. O-Campa{,} and J.D.~García M.
\newblock Electron in bilayer graphene with magnetic fields leading to shape
  invariant potentials.
\newblock {\em Journal of Physics A: Mathematical and Theoretical},
  53(43):435202, 2020.

\bibitem{Bagchi2021}
Bagchi{,} B. and R.~Ghosh.
\newblock {Dirac {H}amiltonian in a supersymmetric framework}.
\newblock {\em Journal of Mathematical Physics}, 62(7):072101, 2021.

\bibitem{Miri2013a}
M.-A. Miri, M.~Heinrich, R.~El-Ganainy, and D.N. Christodoulides.
\newblock {Supersymmetric Optical Structures}.
\newblock {\em Physical Review Letters}, 110(23):233902, 2013.

\bibitem{contreras2019photonic}
A.~Contreras-Astorga{,} and V.~Jakubský.
\newblock Photonic systems with two-dimensional landscapes of complex
  refractive index via time-dependent supersymmetry.
\newblock {\em Physical Review A}, 99(5):053812, 2019.

\bibitem{Chandra2021}
N.~Chandra{,} N.M. and Litchinitser.
\newblock {Photonic bandgap engineering using second-order supersymmetry}.
\newblock {\em Communications Physics}, 4(1), 2021.

\bibitem{Adler1994}
V.E. Adler.
\newblock {Nonlinear chains and Painlev{\'{e}} equations}.
\newblock {\em Physica D: Nonlinear Phenomena}, 73(4):335--351, 1994.

\bibitem{bermudez2011supersymmetric}
D.~Berm{\'u}dez{,} and D.J.~Fern{\'a}ndez C.
\newblock Supersymmetric quantum mechanics and {P}ainlev{\'e} {IV} equation.
\newblock {\em SIGMA. Symmetry, Integrability and Geometry: Methods and
  Applications}, 7:025, 2011.

\bibitem{Correa2016}
F.~Correa{,} A. and Fring.
\newblock {Regularized degenerate multi-solitons}.
\newblock {\em Journal of High Energy Physics 2016 2016:9}, 2016(9):1--16, sep
  2016.

\bibitem{clarkson2020cyclic}
P.~Clarkson{,} D. G{\'o}mez-Ullate{,}~Y. Grandati{,} and R.~Milson.
\newblock Cyclic maya diagrams and rational solutions of higher order
  painlev{\'e} systems.
\newblock {\em Studies in Applied Mathematics}, 144(3):357--385, 2020.

\bibitem{Demircioglu2002}
B.~Demircioglu, Ş~Kuru, M.~{\"{O}}nder, and A.~Ver{\c{c}}in.
\newblock {Two families of superintegrable and isospectral potentials in two
  dimensions}.
\newblock {\em Journal of Mathematical Physics}, 43(5):2133, 2002.

\bibitem{Marquette2009}
I.~Marquette.
\newblock {Superintegrability with third order integrals of motion, cubic
  algebras, and supersymmetric quantum mechanics. I. Rational function
  potentials}.
\newblock {\em Journal of Mathematical Physics}, 50(1):012101, 2009.

\bibitem{adler1994modification}
V.E. Adler.
\newblock A modification of {C}rum's method.
\newblock {\em Theoretical and Mathematical Physics}, 101(3):1381--1386, 1994.

\bibitem{gomez2014extended}
D.~Gomez-Ullate{,}~Y. Grandati{,} and R.~Milson.
\newblock Extended {K}rein-{A}dler theorem for the translationally shape
  invariant potentials.
\newblock {\em Journal of Mathematical Physics}, 55(4):043510, 2014.

\bibitem{Gomez-Ullate2009}
D.~G{\'{o}}mez-Ullate, N.~Kamran, and R.~Milson.
\newblock {An extended class of orthogonal polynomials defined by a
  Sturm–Liouville problem}.
\newblock {\em Journal of Mathematical Analysis and Applications},
  359(1):352--367, 2009.

\bibitem{Quesne2008}
C.~Quesne.
\newblock Exceptional orthogonal polynomials, exactly solvable potentials and
  supersymmetry.
\newblock {\em Journal of Physics A: Mathematical and Theoretical},
  41(39):392001, 2008.

\bibitem{Odake2009}
S.~Odake{,} and R.~Sasaki.
\newblock {Infinitely many shape invariant potentials and new orthogonal
  polynomials}.
\newblock {\em Physics Letters B}, 679(4):414--417, 2009.

\bibitem{marquette2013new}
I.~Marquette{,} and C.~Quesne.
\newblock New ladder operators for a rational extension of the harmonic
  oscillator and superintegrability of some two-dimensional systems.
\newblock {\em Journal of mathematical physics}, 54(10):102102, 2013.

\bibitem{marquette2014combined}
I.~Marquette{,} and C.~Quesne.
\newblock Combined state-adding and state-deleting approaches to type {III}
  multi-step rationally extended potentials: {A}pplications to ladder operators
  and superintegrability.
\newblock {\em Journal of Mathematical Physics}, 55(11):112103, 2014.

\bibitem{marquette2020fourth}
I.~Marquette{,}~S. Post{,} and L.~Ritter.
\newblock A fourth-order superintegrable system with a rational potential
  related to {P}ainlev{\'e} {VI}.
\newblock {\em Journal of Physics A: Mathematical and Theoretical},
  53(50):50LT01, 2020.

\bibitem{marquette2022family}
I.~Marquette{,}~S. Post{,} and L.~Ritter.
\newblock A family of fourth-order superintegrable systems with rational
  potentials related to {P}ainlev{\'e} {VI}.
\newblock {\em Journal of Physics A: Mathematical and Theoretical},
  55(15):155201, 2022.

\bibitem{schrodinger1926stetige}
E.~Schrödinger.
\newblock Der stetige Übergang von der mikro-zur makromechanik.
\newblock {\em naturwissenschaften}, 14(28):664--666, 1926.

\bibitem{glauber1963coherent}
R.J. Glauber.
\newblock Coherent and incoherent states of the radiation field.
\newblock {\em Physical Review}, 131(6):2766, 1963.

\bibitem{barut71}
{A}{.}{O}{.} Barut{,} and L.~Girardello.
\newblock New “coherent” states associated with non-compact groups.
\newblock {\em Communications in Mathematical Physics}, 21(1):41--55, 1971.

\bibitem{perelomov72}
A.M{.} Perelomov.
\newblock Coherent states for arbitrary lie group.
\newblock {\em Communications in Mathematical Physics}, 26(3):222--236, 1972.

\bibitem{nieto79}
M.M. Nieto{,} and L.M. Simmons.
\newblock Coherent states for general potentials. {I}{.} {F}ormalism.
\newblock {\em Phys. Rev. D}, 20:1321--1331, 1979.

\bibitem{dodonov80}
V.V. Dodonov, E.V. Kurmyshev, and V.I. Manko.
\newblock Generalized uncertainty relation and correlated coherent states.
\newblock {\em Physics Letters A}, 79(2):150--152, 1980.

\bibitem{gazeau1999coherent}
J.P. Gazeau{,} and J.R. Klauder.
\newblock Coherent states for systems with discrete and continuous spectrum.
\newblock {\em Journal of Physics A: Mathematical and General}, 32(1):123,
  1999.

\bibitem{hussin1994coherent}
D.J. Fernández C.{,}~V. Hussin{,} and L.M. Nieto.
\newblock Coherent states for isospectral oscillator {H}amiltonians.
\newblock {\em Journal of Physics A: Mathematical and General}, 27(10):3547,
  1994.

\bibitem{Kumar96}
M.S. Kumar{,} and A.~Khare.
\newblock Coherent states for isospectral {H}amiltonians.
\newblock {\em Physics Letters A}, 217:73--77, 1996.

\bibitem{Junker99}
G.~Junker{,} and P.~Roy.
\newblock Non-linear coherent states associated with conditionally exactly
  solvable problems.
\newblock {\em Physics Letters A}, 257(3):113--119, 1999.

\bibitem{Bagchi99}
B.~Bagchi, A.~Ganguly, D.~Bhaumik, and A.~Mitra.
\newblock {Higher derivatives supersymmetry, a modified Crum-Darboux
  transformation and coherent state}.
\newblock {\em Modern Physics Letters A}, 14(1):27--34, 1999.

\bibitem{fernandez2007coherent}
D.J. Fern{\'a}ndez C. {,}~V. Hussin{,} and O.~Rosas-Ortiz.
\newblock Coherent states for {H}amiltonians generated by supersymmetry.
\newblock {\em Journal of Physics A: Mathematical and Theoretical},
  40(24):6491, 2007.

\bibitem{hussin1999higher}
V.~Hussin{,} and D.J.~Fernández C.
\newblock Higher-order {SUSY}, linearized nonlinear {H}eisenberg algebras and
  coherent states.
\newblock {\em Journal of Physics A: Mathematical and General}, 32(19):3603,
  1999.

\bibitem{Bermudez2014}
D.~Bermudez, A.~Contreras-Astorga, and D.J. {Fern{\'{a}}ndez C.}
\newblock {Painlev{\'{e}} IV coherent states}.
\newblock {\em Annals of Physics}, 350:615--634, 2014.

\bibitem{hussin2017coherent}
V.~Hussin{,} D.J.~Fernández C.{,} and V.S. Morales-Salgado.
\newblock Coherent states for supersymmetric partners of the infinite well.
\newblock In {\em Journal of Physics. Conference Series}, volume 839, 2017.

\bibitem{hoffmann2018coherent}
S.E. Hoffmann{,} V. Hussin{,}~I. Marquette{,} and Z.~Yao-Zhong.
\newblock Coherent states for ladder operators of general order related to
  exceptional orthogonal polynomials.
\newblock {\em Journal of Physics A: Mathematical and Theoretical},
  51(31):315203, 2018.

\bibitem{hoffmann2018non}
S.E. Hoffmann{,} V. Hussin{,}~I. Marquette{,} and Z.~Yao-Zhong.
\newblock Non-classical behaviour of coherent states for systems constructed
  using exceptional orthogonal polynomials.
\newblock {\em Journal of Physics A: Mathematical and Theoretical},
  51(8):085202, 2018.

\bibitem{hussin2019coherent}
D.J. Fernández C.{,}~V. Hussin{,} and V.S. Morales-Salgado.
\newblock Coherent states for the supersymmetric partners of the truncated
  oscillator.
\newblock {\em The European Physical Journal Plus}, 134(1):1--15, 2019.

\bibitem{hoffmann2019ladder}
S.E. Hoffmann{,} V. Hussin{,}~I. Marquette{,} and Z.~Yao-Zhong.
\newblock Ladder operators and coherent states for multi-step supersymmetric
  rational extensions of the truncated oscillator.
\newblock {\em Journal of Mathematical Physics}, 60(5):052105, 2019.

\bibitem{garneau2021ladder}
S.~Garneau-Desroches{,} and V.~Hussin.
\newblock Ladder operators and coherent states for the {R}osen--{M}orse system
  and its rational extensions.
\newblock {\em Journal of Physics A: Mathematical and Theoretical},
  54(47):475201, 2021.

\bibitem{quesne2011higher}
C.~Quesne.
\newblock Higher-order susy, exactly solvable potentials, and exceptional
  orthogonal polynomials.
\newblock {\em Modern Physics Letters A}, 26(25):1843--1852, 2011.

\bibitem{witten1981dynamical}
E.~Witten.
\newblock Dynamical breaking of supersymmetry.
\newblock {\em Nuclear Physics B}, 188(3):513--554, 1981.

\bibitem{carballo2004polynomial}
J.M Carballo{,} D.J. Fernández C.{,}~J. Negro{,} and L.M. Nieto.
\newblock Polynomial {H}eisenberg algebras.
\newblock {\em Journal of Physics A: Mathematical and General}, 37(43):10349,
  2004.

\bibitem{odake2013krein}
S.~Odake{,} and R.~Sasaki.
\newblock Krein-{A}dler transformations for shape-invariant potentials and
  pseudo virtual states.
\newblock {\em Journal of Physics A: Mathematical and Theoretical},
  46(24):245201, 2013.

\bibitem{gomez2013rational}
D.~G{\'o}mez-Ullate{,}~Y. Grandati{,} and R.~Milson.
\newblock Rational extensions of the quantum harmonic oscillator and
  exceptional {H}ermite polynomials.
\newblock {\em Journal of Physics A: Mathematical and Theoretical},
  47(1):015203, 2013.

\bibitem{mielnik1984factorization}
B.~Mielnik.
\newblock Factorization method and new potentials with the oscillator spectrum.
\newblock {\em Journal of mathematical physics}, 25(12):3387--3389, 1984.

\bibitem{weisner1959generating}
L.~Weisner.
\newblock Generating functions for {H}ermite functions.
\newblock {\em Canadian Journal of Mathematics}, 11:141--147, 1959.

\bibitem{abramowitz}
M.~{Abramowitz}{,} and I.A. {Stegun}.
\newblock {\em Handbook of Mathematical Functions with Formulas, Graphs, and
  Mathematical Tables}.
\newblock Dover, New York, 1964.

\bibitem{erdelyi1953higher}
A.~Erd{\'e}lyi.
\newblock {\em Higher transcendental functions Vol. I}.
\newblock McGraw-Hill, New York, 1953.

\bibitem{PhysRev.40.749}
E.~Wigner.
\newblock On the quantum correction for thermodynamic equilibrium.
\newblock {\em Phys. Rev.}, 40:749--759, 1932.

\bibitem{mandel1979sub}
L.~Mandel.
\newblock Sub-poissonian photon statistics in resonance fluorescence.
\newblock {\em Optics letters}, 4(7):205--207, 1979.

\end{thebibliography}

\end{sloppypar}
\end{document}